\documentclass[structabstract]{aa}
\usepackage{epsfig}

\usepackage{graphicx} 
\usepackage{array}

\def\aap{A\&A}                

\usepackage{tabularx}
\usepackage{longtable}
\usepackage{hyperref}

\usepackage{natbib,twoopt}

\begin{document}

\newcommand{\kms}{km~s$^{-1}$}
\newcommand{\Msun}{M$_{\odot}$}
\newcommand{\Teff}{$T_{\rm eff}$}
\newcommand{\FeH}{[Fe/H]}
\newcommand{\Cir}{$^{12}$C / $^{13}$C}
\newcommand{\bacchus}{{\footnotesize BACCHUS}}
\newcommand{\vald}{{\footnotesize VALD}}

\title{Hertzsprung-Russell diagram and mass distribution \\
of  barium stars
\thanks{Table A.1 is only available in electronic form at the CDS via anonymous ftp to cdsarc.u-strasbg.fr (130.79.128.5) or via \url{http://cdsweb.u-strasbg.fr/cgi-bin/qcat?J/A+A/}}
}
\author{
A. Escorza\inst{1,2}
\and 
H.M.J. Boffin\inst{3}
\and
A.Jorissen\inst{2}
\and
S. Van Eck\inst{2}
\and 
L. Siess\inst{2}
\and
H. Van Winckel\inst{1}
\and
D. Karinkuzhi\inst{2}
\and
\\
S. Shetye\inst{2,1}
\and
D. Pourbaix\inst{2}
}
\institute{Institute of Astronomy, KU Leuven, Celestijnenlaan  200D, 3001 Leuven, Belgium
\and
Institut d'Astronomie et d'Astrophysique, Universit\'e Libre de Bruxelles, ULB, Campus Plaine C.P. 226, Boulevard du Triomphe, B-1050 Bruxelles, Belgium
\and
ESO, Karl Schwarzschild Stra\ss e 2, D-85748 Garching bei M\"unchen, Germany
}

\date{Received; 
Accepted}

\abstract{With the availability of parallaxes provided by the Tycho-Gaia Astrometric Solution, it is possible to construct the Hertzsprung-Russell diagram (HRD) of barium and related stars with unprecedented accuracy. A direct result from the derived HRD is that subgiant CH stars occupy the same region as barium dwarfs, contrary to what their designations imply. By comparing the position of barium stars in the HRD with STAREVOL evolutionary tracks, it is possible to evaluate their masses, provided the metallicity is known. We used an average metallicity [Fe/H] = $-$0.25 and derived the mass distribution of barium giants. The distribution peaks around 2.5 M$_\odot$ with a tail at higher masses up to 4.5~M$_\odot$. This peak is also seen in the mass distribution of a sample of normal K and M giants used for comparison and is associated with stars located in the red clump. When we compare these mass distributions, we see a deficit of low-mass (1 -- 2 M$_\odot$) barium giants. This is probably because low-mass stars reach large radii at the tip of the red giant branch, which may have resulted in an early binary interaction. Among barium giants, the high-mass tail is however dominated by stars with barium indices of less than unity, based on a visual inspection of the barium spectral line; that is, these stars have a very moderate barium line strength. We believe that these stars are not genuine barium giants, but rather bright giants, or supergiants, where the barium lines are strengthened because of a positive luminosity effect. Moreover, contrary to previous claims, we do not see differences between the mass distributions of mild and strong barium giants.
 
} 


\maketitle

\section{Introduction}
Barium (Ba) giants are a class of G- and K-type giants with strong spectral lines of elements, such as barium or strontium, produced by the slow neutron capture (s-) process of nucleosynthesis \citep{Kappeler2011}. Although the class was already defined in 1951 \citep{Bidelman1951}, it was not until 1980 \citep{McClure1980} that the origin of the observed overabundances was understood as the result of mass transfer in a binary system. The polluting heavy elements were formerly produced within an asymptotic giant branch (AGB) companion, which is now a very faint white dwarf (WD) that cannot be directly observed in most cases.

Although they have been less intensively studied, barium stars are also found in the main sequence (i.e. barium dwarfs, dBa; \citealt{JorissenBoffin92}; \citealt{North00}). Additionally, closely related to Ba stars, CH stars are their low metallicity counterparts. The latter have similar enhancement of s-process elements and strong CH molecular bands, but weaker lines of other metals \citep{Keenan1942}.

The exact mode of mass transfer responsible for the formation of these families of polluted binaries remains uncertain. Many systems have orbital periods in a range (form 100 to 1000~days) that cannot be accounted for by simple models of orbital evolution (\citealt{Pols2003}; \citealt{Izzard2010}, and references therein). Nevertheless, the white-dwarf nature of the companion is beyond doubt, thanks to the analysis of the orbital mass functions performed by \citet{Webbink1988}, \citet{McClure1990}, and \citet{VanderSwaelmen2017}. The latter analysis reveals a mass distribution for the WD companion which peaks around 0.6~M$_\odot$, in accordance with the expectation. However, the exact value of the derived WD mass depends upon the mass of the primary star (the barium star), which is difficult to derive with a good accuracy.

The mass of barium stars may be derived from their location in the Hertzsprung-Russell diagram (HRD) and from a comparison with evolutionary tracks. \citet{Mennessier1997} used a Bayesian method to infer barium star masses, based on a HRD constructed  from  Hipparcos  parallaxes. These authors  concluded  that  mild  and  strong barium stars have somewhat different mass distributions, which are characterized by masses in the range 2.5 -- 4.5~M$_\odot$ and 1 -- 3~M$_\odot$, respectively. The distinction between mild and strong barium stars is made on the \textit{Ba index} introduced by \citet{Warner1965}. The index reflects the strength of the barium spectral lines, based on visual inspection, on a scale from Ba1 to Ba5, where Ba5 corresponds to the strongest lines. In this and our past studies, we associate Ba1 - Ba2 indices with mild  barium  stars  and  Ba3  -  Ba5  indices  with  strong  barium stars. The catalogue of barium stars by \citet{Lu1991} introduces many barium stars with a \textit{Ba index} smaller than one, which we denote as Ba0. These targets deserve special attention as they may turn out not to be barium stars.

\begin{figure}
\includegraphics[width=0.50\textwidth]{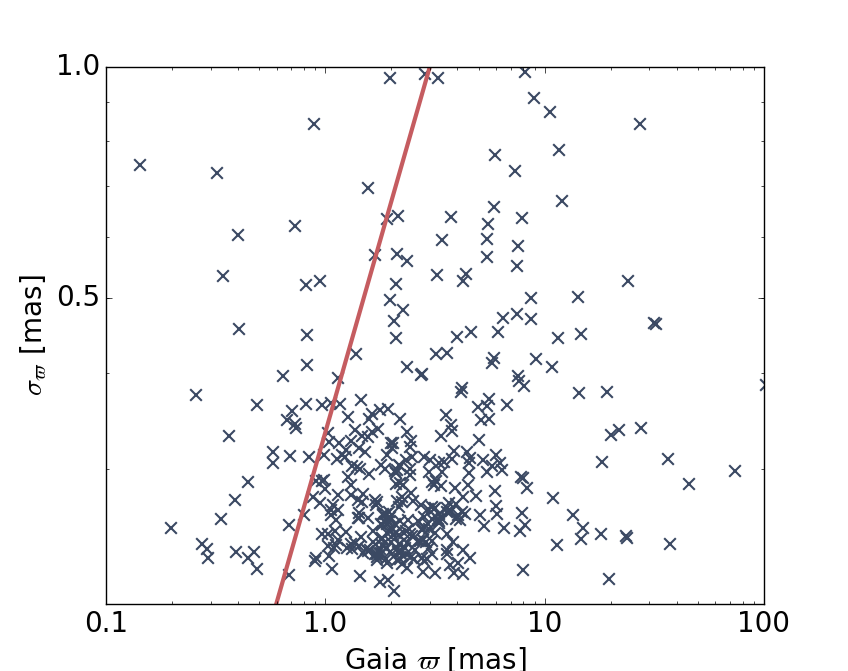}
\caption{\label{Fig:varpi-errpi}
Parallax and its error for the sample of Ba and related stars studied in this paper that were part of TGAS. The red line represents the threshold $\varpi /\sigma(\varpi) = 3$. Only stars to the right of the line are considered for inclusion in the HRD. 
}
\end{figure}

Thanks to the recent data release (DR1; \citealt{Gaia-collaborationDR1}) of the more accurate parallaxes provided by the Tycho-Gaia Astrometric Solution (TGAS; \citealt{Michalik2015}; \citealt{Lindegren16}), it is possible to locate barium stars in the HRD with a much better accuracy than with the Hipparcos data. Stellar models still suffer from large uncertainties, but the Gaia mission \citep{GaiaMission} will mark a significant step forward for the derivation of barium star masses.

The relation between the location of barium stars in the HRD and the orbital parameters also provides valuable insights into the binary interaction processes at play in these systems. For instance, one may expect that for barium stars located in the red clump, the distribution of orbital periods exhibits a higher lower limit caused by the large radii reached by the star at the tip of the red giant branch (RGB). One may also expect that short-period systems for which the red giant has gone through the RGB tip, show smaller or null eccentricities due to tidal circularization along the RGB.

The paper is organized as follows: The sample is described in Sect.~\ref{Sect:TGAS} and the HRD of barium stars and related systems is then constructed in Sect.~\ref{Sect:HRD}. We discuss the resulting mass distribution of barium giants in Sect.~\ref{Sect:masses} and we investigate the possible relationship between location in the HRD and orbital period in Sect.~\ref{Sect:P-HRD}. \nocite{North1994,Luck-Bond-1991,Keenan1942} 

\section{The sample}
\label{Sect:TGAS}

The sample was constructed by selecting targets with Tycho-2 identifiers \citep{Hog2000} in the barium star lists of \citet{Lu1983} and \citet{Lu1991}, in the list of CH and related stars of \citet{Bartkevicius1996}, and in the list of dwarf barium stars of \citet{North1994} and \citet{Edvardsson1993}. The sample of (dwarf and giant) barium, CH and related stars with a Tycho-2 identifier amounts to 546 entries. The TGAS catalogue \citep{Lindegren16} provides a parallax value for 400 of them. We removed from the  \citet{Bartkevicius1996} list 11 high proper motion dwarf stars, labelled PM* in the SIMBAD database \citep{SIMBAD}, for which we could not find any confirmation that they are either carbon or barium stars (e.g. HD 208998, neither a carbon star nor a Ba star according to \citealt{Bensby2006,Bond2008}). The list of stars removed for that reason includes HD 11397, HD 15206, HD 24508, HD 89668, HD 108564, HD 145417, HD 153075, HD 154276, HD 161612, HD 164922, and HD 208998.

\begin{figure}
\includegraphics[width=0.49\textwidth]{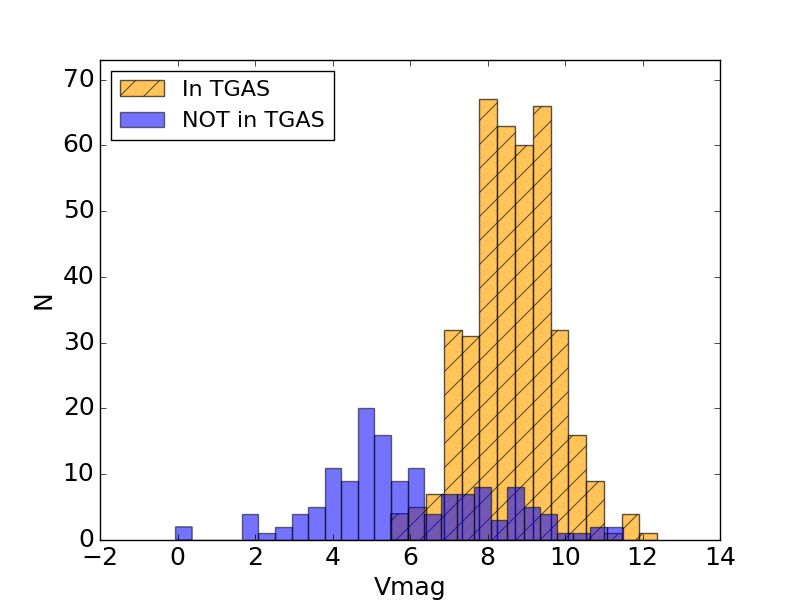}
\caption{\label{Fig:Vhisto}
$V$ magnitude distribution of our stellar sample according to the presence or absence of the star in the TGAS catalogue.
}
\end{figure}

Fig.~\ref{Fig:varpi-errpi} shows in logarithm scale the relation between the parallax ($\varpi$) and its error ($\sigma(\varpi)$) for the TGAS targets. The red line represents the limit where $\varpi /\sigma(\varpi) \ge 3$, which was used as precision condition. Among the initial 389 barium and related stars with a TGAS parallax, after removing the 11 high proper motion stars, only 52 (13\%) had to be rejected because they do not fulfil the precision criterion. Since this rejection rate is small, and since the analysis of these data do not significantly affect any statistical property of the sample, such as its average luminosity, there is no need to investigate how this rejection rate could bias the results. Nine more targets were excluded because their position within the Galaxy made it difficult to constrain the interstellar extinction on the line of sight, which, as explained below, prevents a reliable derivation of the stellar effective temperature. 

When the TGAS parallax was not available, we used the Hipparcos value \citep{ESA1997}, always imposing the condition that $\varpi/\sigma(\varpi)$ exceeds 3. Additionally, for 21 confirmed astrometric binaries, we used the parallax rederived by \cite{Pourbaix2000} instead of the TGAS (available for 15 of them) or Hipparcos values. The reason to do this for the astrometric binaries is that the TGAS parallax, obtained by applying a single star solution, is possibly disturbed by the orbital motion. The possible presence of astrometric binaries in the sample is discussed at the end of the present section. This results in 103 entries from Hipparcos and 313 from TGAS, in addition to the 21 astrometric binaries. Our final sample thus contains 437 entries. Of these, 77\% are barium giants while the other samples are much smaller; i.e. 10\% of the stars are barium dwarfs, 5\% are CH giants, 5\% are carbon stars, and the remaining 3\% are CH subgiants.

\begin{figure}
\includegraphics[width=0.49\textwidth]{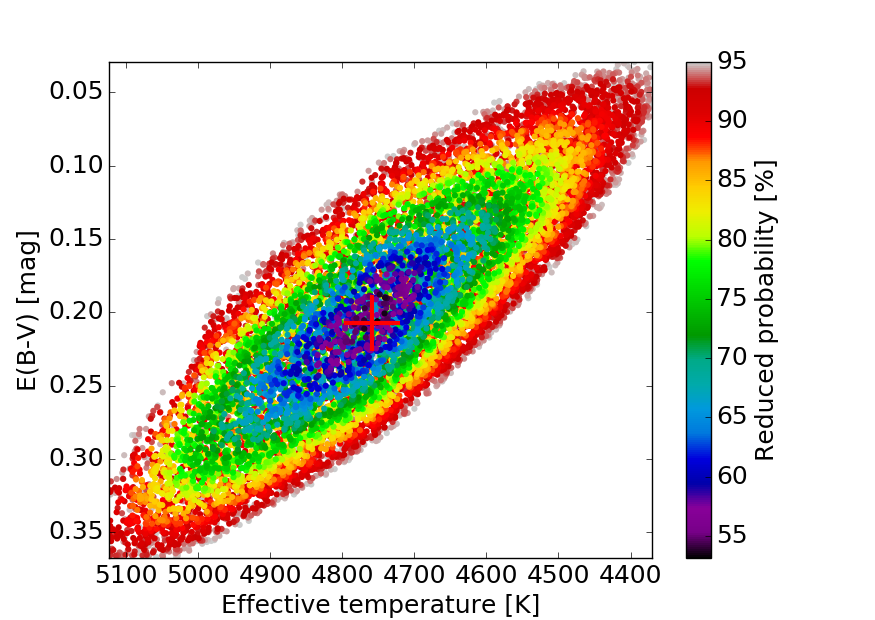}
\caption{\label{Fig:EB-V_fit}
Error ellipse resulting from fitting the SED of HD 183915, revealing the strong correlation between $T_{\rm eff}$ and $E_{\rm B-V}$. The red cross identifies the best fit, predicting $E_{\rm B-V} = 0.203$ whereas  the selective extinction derived from \citet{Gontcharov2012} for that star is $E_{\rm B-V} = 0.225$ (leading to $A_V = 0.70$). 
}
\end{figure}

As shown in Fig.~\ref{Fig:Vhisto}, the major cause for a star not to be included in the TGAS catalogue is because it is brighter than $V \sim 6.5$. A small fraction of fainter stars are missing as well. In an attempt to identify whether their absence in TGAS could be related to their astrometric binary nature, stars with a missing TGAS parallax and a known orbital period (see Jorissen et al. 2017, in preparation) were collected in Table~\ref{Tab:missing}. In that Table, stars are ordered by increasing  $a_1/\varpi$ (seventh column), the ratio between the angular semi-major axis of the astrometric orbit of the primary component around the centre of mass of the system and the parallax. This ratio may be estimated from Eq.~(13) of \citet{Pourbaix2000}, which depends only on the masses and the orbital period

\begin{equation}\label{Eq:a1varpi}
\frac{a_1}{\varpi} = P^{2/3} \frac{M_2}{(M_1+M_2)^{2/3}},
\end{equation}

\noindent where $a_1/\varpi$ is in AU, $P$ is expressed in years and masses in \Msun. 
This quantity corresponds in fact to the semi-major axis of the absolute orbit of the primary component. In case the astrometric orbit was detected from the Hipparcos data \citep{Pourbaix2000}, the values listed in the seventh column of Table~\ref{Tab:missing} are the observed values, which correspond to the orbit of the photocentre of the system around its centre of mass. However, since the cases considered here correspond to a WD companion, the photocentric orbit is identical to that of the primary component. If the astrometric orbit is not seen in the Hipparcos data, the ratio $a_1/\varpi$ is estimated from Eq.~\ref{Eq:a1varpi}
by adopting $M_1 =  2.5$~\Msun\  and $M_2 = 0.62$~\Msun, along with the observed orbital period.

A high $a_1/\varpi$ value means that the binary motion with respect to the parallactic motion is large. If, in addition, the orbital period is close to 1~yr, the parallactic and binary motions are difficult to disentangle \citep{Pourbaix2000}. This could be a possible cause for the absence of a star in the TGAS catalogue. However, this is not observed because the fainter stars with high $a_1/\varpi$ values absent from TGAS often have orbital periods much longer than 1~yr. Moreover, several systems with similarly high $a_1/\varpi$ ratios \citep{Pourbaix2000} do have TGAS parallaxes (like HD~50264, HD~87080, HD~107574...). Another possible cause for the absence of a star in the TGAS catalogue could be its colour, as very red and very blue stars were excluded \citep{Gaia-collaborationDR1}. The identification of the exact reason for the absence of several faint barium and related stars has to await the availability of the next data releases when quality flags of the astrometric solution become available.

\begin{figure}
\centering
\includegraphics[width=0.49\textwidth]{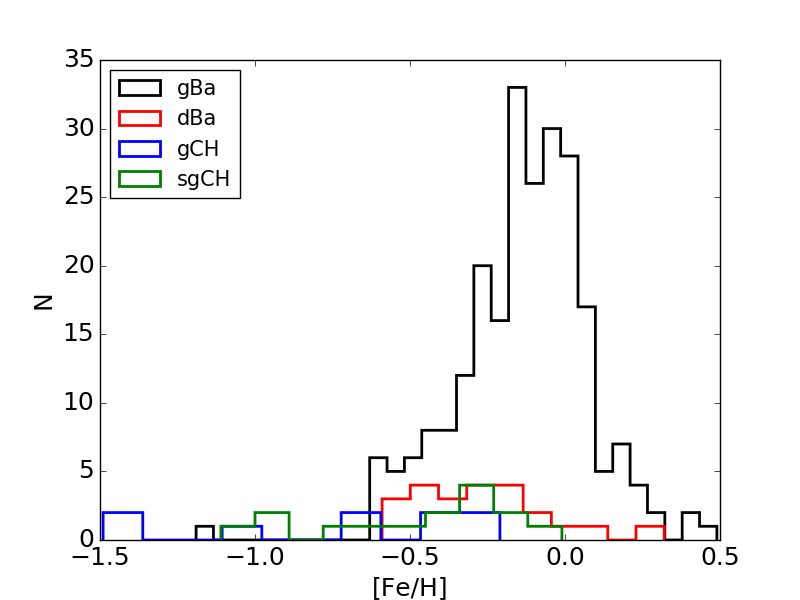}
\caption{\label{Fig:FeH}
Metallicity distribution of the barium and related stars with a value of [Fe/H] available in the literature.
}
\end{figure}

\section{Hertzsprung-Russell diagram}
\label{Sect:HRD}

\subsection{Atmospheric parameters}\label{Sect:atmosphere}
\label{Sect:params}

\begin{table*}
\caption[]{\label{Tab:missing}
Barium stars not listed in the TGAS catalogue, and with a known orbital period ($P$), ordered by increasing $a_1/\varpi$ values. $G$ is the Gaia magnitude and $Ks$ is from the 2MASS survey \citep{2MASS}, $\varpi$ is from the Hipparcos catalogue \citep{ESA1997} or from \citet{Pourbaix2000} when available (PJ in column 'Rem.'). An asterisk in the 'Rem.' column means that the Hipparcos parallax was used to locate the star in the HRD. Column $a_1/\varpi$ is obtained from \citet{Pourbaix2000} or from Eq.~\ref{Eq:a1varpi} for astrometric orbits not detected in the Hipparcos data.
}
\begin{tabular}{lllllllllllll}
\hline
HD/BD & TYC &  $V$ & $G$ & $G-Ks$ & $\varpi$(HIP) (mas) &  $a_1/\varpi$ (AU) & $P$ (d) & 
Rem \\
\hline\\
\noalign{Ba strong}
\medskip\\
121447  & 6140-641-1 & 7.80 & 7.13 & 2.98 & $ 2.2\pm1.0$ & 0.2 & 186 \\
46407 & 5369-220-1   &  6.24 & - &  - & 6.6${+1.3}\atop{-1.1}$ & 0.3 & 457 &  PJ,*\\
92626 & 8201-1209-1 & 7.09 & 6.67 & 2.25 & 6.6${+0.9}\atop{-0.6}$ & 0.5 & 918 &  PJ,*\\
NGC 2420 250 & 1373-1426-1 & 11.14 & 11.01 & 2.37& - & 0.7 & 1404 \\
101013 &  3454-2188-1 & 6.12 & - & -  & 7.1${+0.7}\atop{-0.6}$ & 0.8 & 1711 &  PJ,*\\
+38 118 & 2797-46-1  & 8.86 & 8.32 & 2.76  & - & 1.4 & 3877 \\
\medskip\\
\noalign{Ba mild}
\medskip\\
77247 & 3805-1493-1 & 6.86 & 6.54 & 1.89 & $2.9\pm1.0$ & 0.1 & 80 & *\\
218356 & 2239-1475-1 & 4.54 & - & - &  $6.1\pm0.7$ & 0.1 & 111 & *\\
288174 & 119-1058-1 & 9.02 & - & - & $2.9\pm1.3$ & 0.8 & 1818\\
204075 & 6372-1278-1 & 3.74 & - & - & 8.6${+1.1}\atop{-1.0}$& 1.0 &  2378 &  PJ,*\\
131670 & 4999-334-1 & 8.01 & 7.62 & 2.31 & $2.3\pm1.2$ & 1.2 & 2930 \\
139195 & 933-1240-1 & 5.26 & - & - & $13.9\pm0.7$ &1.7 & 5324 &*\\
53199 & 761-980-1 & 9.07 & 8.79 & 1.79 & $3.7\pm1.3$ & 2.3 & 8300 &*\\
51959 & 4813-1015-1 & 8.92 & 8.61 & 2.07 & $6.5\pm1.3$ & 2.5 & 9488  &*\\
104979 & 866-1180-1 & 4.12 &  - & - & $19.1\pm0.8$  & 3.3 & 13940  &*\\
98839 & 3015-2321-1 & 5.03 & - & - &  $6.6\pm0.6$ & 3.7 & 16419 & * \\
119185 & 5552-1079-1 & 8.91 & 8.57 & 2.02 & $3.9\pm1.1$ &4.1 & 19467 & *\\
\medskip\\
\noalign{Ba dwarf}
\medskip\\
89948 & 6631-715-1   & 7.55 & 7.31 & 1.12 & $23.4\pm0.9$ & 0.7 & 667.8 &  PJ,*\\
76225 & 6580-2586-1 & 9.20 &  - & - & $3.4\pm1.1$ &1.0 & 2411 & * \\
98991 & 6088-2156-1 & 5.09 & - & - & $22.0\pm0.8$ &1.1 & 2834 & * \\
221531 & 5832-970-1 & 8.36 & 8.21 & 0.98 &9.6${+1.4}\atop{-1.3}$  & 1.2  & 1416 &  PJ,*\\
95241 & 3012-2522-1 & 6.03 & - & - & $22.0\pm0.8$ &1.8 & 5448 & *\\
\hline\\
\end{tabular}
\end{table*}

\begin{figure}
\includegraphics[width=0.49\textwidth]{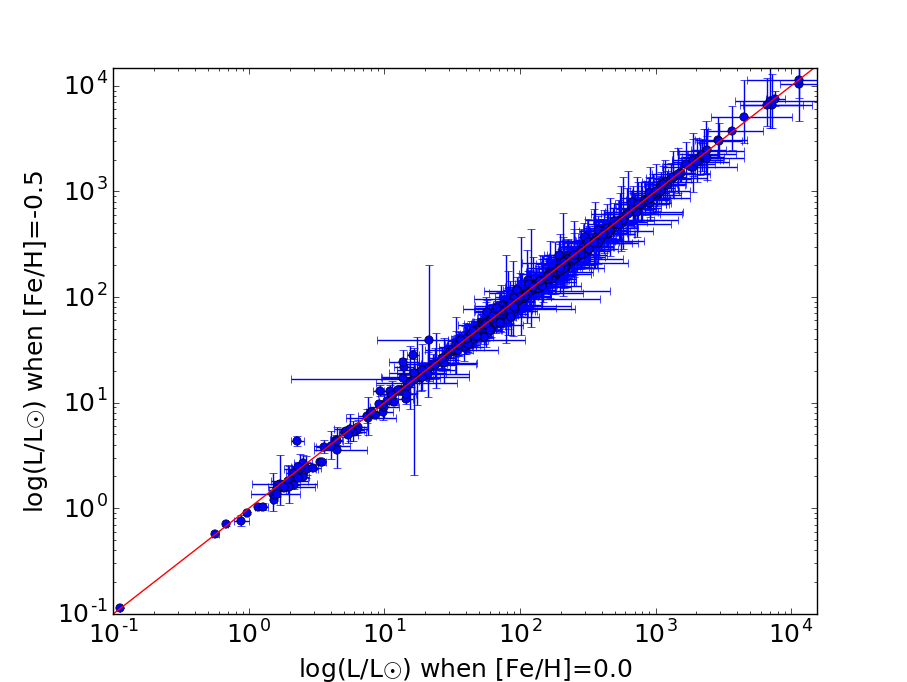}
\includegraphics[width=0.49\textwidth]{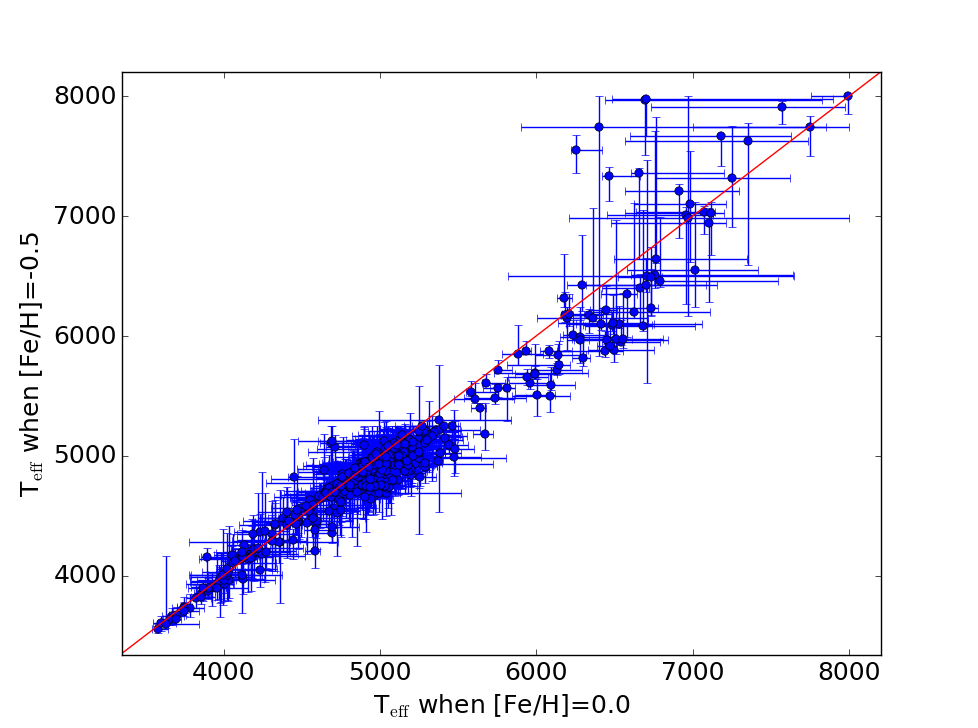}
\caption{\label{Fig:LandT}
Luminosity and \Teff (and their 1$\sigma$ error bar) derived from SED fitting with models of solar metallicity and metallicity [Fe/H]$ = -0.5$.
}
\end{figure}

\begin{figure}
\includegraphics[width=0.49\textwidth]{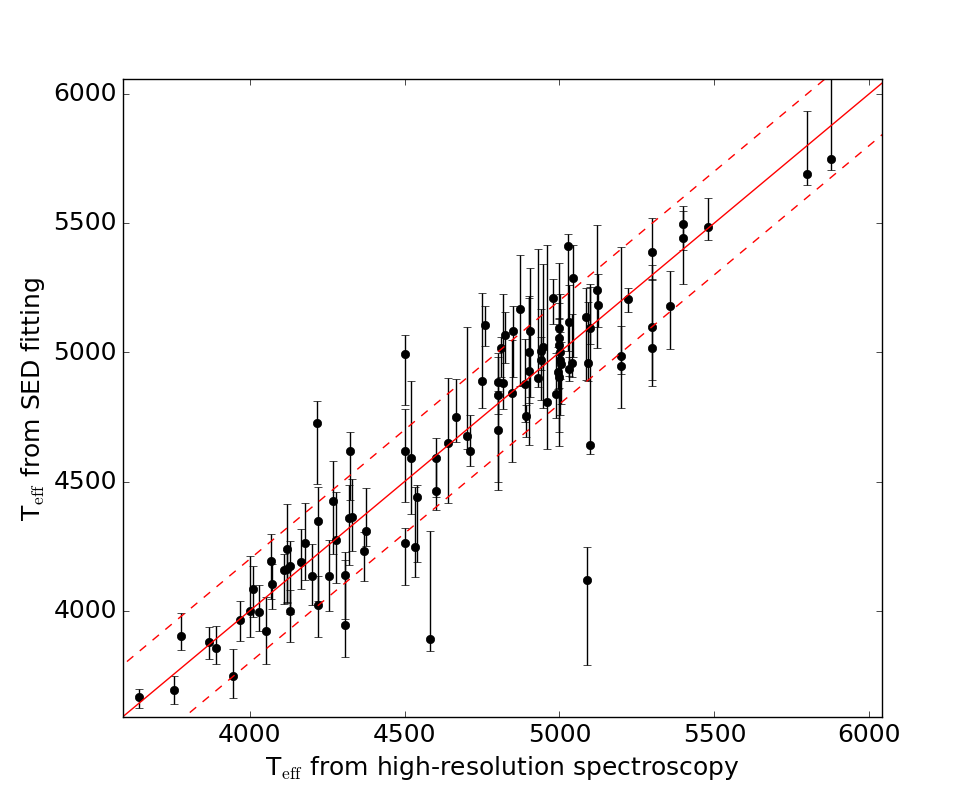}
\caption{\label{Fig:Pastel}
Effective temperatures obtained with the SED fitting method compared with those obtained from high-resolution spectra collected from the PASTEL catalogue \citep{PASTEL16}. The dashed red lines indicate a $\pm$\,200 K uncertainty.
}
\end{figure}

The atmospheric parameters of barium and related stars were derived by modelling the spectral energy distribution (SED) obtained by collecting  magnitudes listed in the Simbad database  \citep{SIMBAD}. The best-fitting MARCS model \citep{Gustafsson2008} was determined from a parameter-grid search using a $\chi^2$ minimization method (see \citealt{Degroote2011} for details). The stellar temperature was then assigned from the best-fitting model and the luminosity was obtained by integrating the SED over all wavelengths spanned by the model, and applying the distance modulus derived from the parallax. The error bar on the luminosity is propagated from the parallax uncertainty and is thus asymmetric. The error on the temperature is the 1-sigma error enclosing 67\% of the model fits (see Fig.~\ref{Fig:EB-V_fit}).

The reddening $E_{\rm B-V}$ was initially left free during the fitting process. Its best value was estimated by looking for the amount of reddening to be applied to the MARCS models to match the observed magnitudes. A value of  3.1 was used for the ratio of the total to the selective extinction $R = A_V/E_{\rm B-V}$ \citep{Weingartner2001}, from which $A_V$ is derived. It turned out that the temperature and extinction derived in this manner are strongly correlated, as shown in Fig.~\ref{Fig:EB-V_fit}.

In a second run, the extinction was fixed at the value computed by \citet{Gontcharov2012}, in his three-dimensional map of the extinction within the nearest kiloparsec. The location of the target in the Galaxy was computed from its galactic coordinates and its parallax. The resulting \Teff\ and $E_{\rm B-V}$ values were often found to fall at the lower edge of the 1$\sigma$ ellipse error. We decided to let $E_{\rm B-V}$ vary freely between 0 and Gontcharov's value plus 0.07~mag, which is the 1$\sigma$ error bar suggested in their paper.

Another parameter that has an impact on the determination of the stellar parameters is the metallicity. As they are difficult to constrain by fitting SEDs, metallicities of barium and related stars were collected from the literature and their distribution is presented in Fig.~\ref{Fig:FeH}. The barium-giant distribution peaks at [Fe/H] $\sim -$0.14 with a standard deviation of 0.2. \citet{deCastro16} determined the metallicity of 182 Ba giants and obtained a similar mean value of [Fe/H]\,$\sim -$\,0.12\,$\pm$\,0.14. As expected, CH stars have lower metallicities, and it seems that barium dwarfs may be slightly more metal poor than barium giants as well, but the dwarf sample is much smaller. We decided to use MARCS models with fixed [Fe/H] = $-$0.25 to determine the stellar parameters. To evaluate the impact of the metallicity on the derived parameters, two more SED fits have been carried out: one imposing a solar metallicity and another one imposing [Fe/H] = $-$0.5. The resulting \Teff\ and luminosities of the limiting cases are presented in Fig.~\ref{Fig:LandT}, and reveal that the uncertainty introduced on the stellar parameters by the metallicity of the best-fitting MARCS model is negligible. The impact of the metallicity on the stellar tracks and hence on the derived mass is of course much larger and is discussed in Sect.~\ref{Sect:masses}. 

Finally, in order to evaluate our method, we include (Fig.~\ref{Fig:Pastel}) a comparison of the effective temperatures, which we obtained for the Ba giants with the SED fitting method with temperatures for the same stars determined from high-resolution spectra. We collected the latter values from the PASTEL catalogue \citep{PASTEL16}. Fig.~\ref{Fig:Pastel} shows that most of the 102 Ba giants that we compare lie within the $\pm$\,200\,K uncertainty, indicated by the dashed red lines. All the parameters described so far are listed in Table~\ref{Tab:master}.

\subsection{Input physics for the stellar grid calculations} 
\label{Sect:STAREVOL}
To determine the fundamental parameters of our sample stars from their location in the HR diagram, we computed extended grids of stellar models covering the mass range $0.6 \le M_\mathrm{zams}/$M$_\odot \le 6$. All the models were evolved from the pre-main sequence up to the end of the AGB phase for stars of initial mass $M_\mathrm{zams} \le 4 $M$_\odot$ and up to the occurrence of convergence problems in the higher mass tracks. The models were computed with the {\sf STAREVOL} code \citep{Siess00,SiessArnould08} with the following input physics. We used the radiative opacity tables of \citet{IglesiasRogers96} above 8000~K and of \citet{Ferguson05} at lower temperatures. We took the modifications of the opacity due to the formation of H2, H2O, OH, C2, CN, and CO  molecules in the atmosphere of C-rich stars (C/O>1) into account following the formulation of \citet{Marigo02}. The nuclear  network includes 182 reactions coupling 55 species from H to Cl. The nuclear rates have been recently updated with NACRE II compilation. We used the mixing length theory to determine the temperature gradient in the convective regions with $\alpha_\mathrm{MLT} = 1.75$ and adopt the Schwarzschild criterion. For the mass loss rate, we considered the \citet{SchroderCuntz07} prescription up to the beginning of the AGB phase and then switched to the \citet{VassiliadisWood93} formulation.  We also included some overshooting at the base of the convective envelope, following the exponential decay expression of \citet{Herwig97}  with $f_\mathrm{over} = 0.1$. Finally we used a grey atmosphere surface boundary condition.

\subsection{Discussion}
\label{Sect:HRDdisc}

\begin{figure*}
\begin{center}
\includegraphics[width=0.89\textwidth]{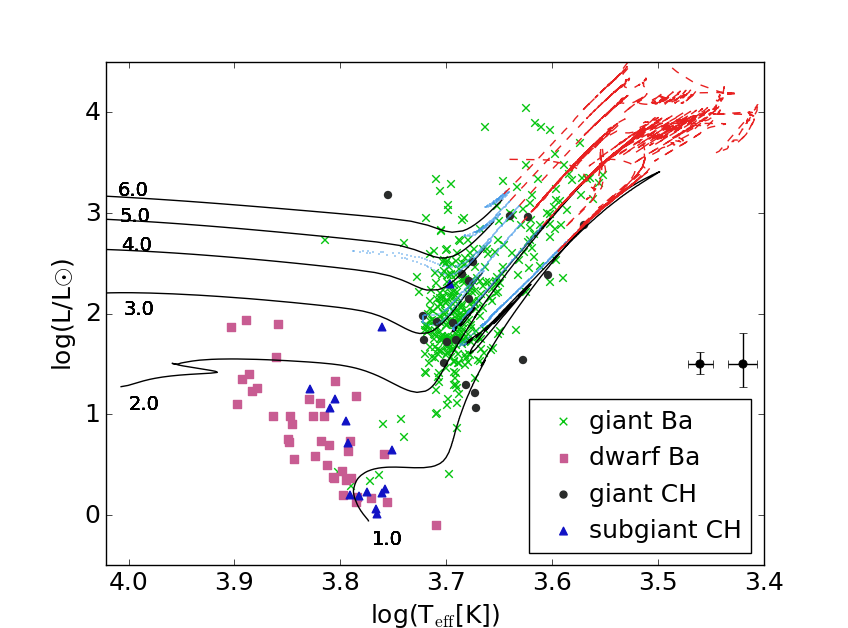}
\caption{\label{Fig:HRDStarevol}
Hertzsprung-Russell diagram for sample of barium and related stars labelled as in queried catalogues. Green crosses indicate barium
giants; pink squares indicate barium dwarfs; black dots indicate CH giants; and blue triangles indicate subgiant CH stars. Typical error bars are shown over the legend for two situations: a favourable case (left: $\varpi /\sigma(\varpi) = 8$), and limiting case for $\varpi /\sigma(\varpi) = 3$ (right). Stellar tracks from the STAREVOL code \citep{Siess2006} were overplotted for models of masses 1.0, 2.0, 3.0, 4.0, 5.0, and 6.0~\Msun, as labelled and metallicity [Fe/H] = $-$0.25. The solid black tracks correspond to the post-main-sequence evolution up to the tip of the RGB (low-mass stars) or to the onset of core He-burning (intermediate-mass stars), dotted blue tracks correspond to core He-burning, and dashed red tracks to the early and thermally pulsing AGB (TP-AGB) phase.
}
\end{center}
\end{figure*}

Fig.~\ref{Fig:HRDStarevol} shows the Ba and related star sample in the HRD with parameters obtained as described in Sect.~\ref{Sect:atmosphere}, i.e. leaving the extinction free (between 0. and Gontcharov's value), fixing the metallicity of the atmospheric models to [Fe/H] = $-$0.25, and by restricting the sample to stars with $\varpi /\sigma(\varpi) \ge 3$. We kept the original category (Ba giant, Ba dwarf, CH giant, or CH subgiant) given in the queried catalogues for each star shown in Fig.~\ref{Fig:HRDStarevol}. Typical error bars are shown for two situations: a favourable case with $\varpi /\sigma(\varpi) = 8$ (left) and limiting case with $\varpi /\sigma(\varpi) = 3$ (right). In the latter case, the asymmetric nature of the errors on the luminosity starts to be noticeable, so that biases are manifested (e.g. \citealt{Luri-1997}). However, as apparent on Fig.~\ref{Fig:varpi-errpi}, the number of targets with $\varpi /\sigma(\varpi) < 3$ is not large, and a discussion on the effect of the biases does not seem to be required here.

\begin{figure}
\includegraphics[width=0.49\textwidth]{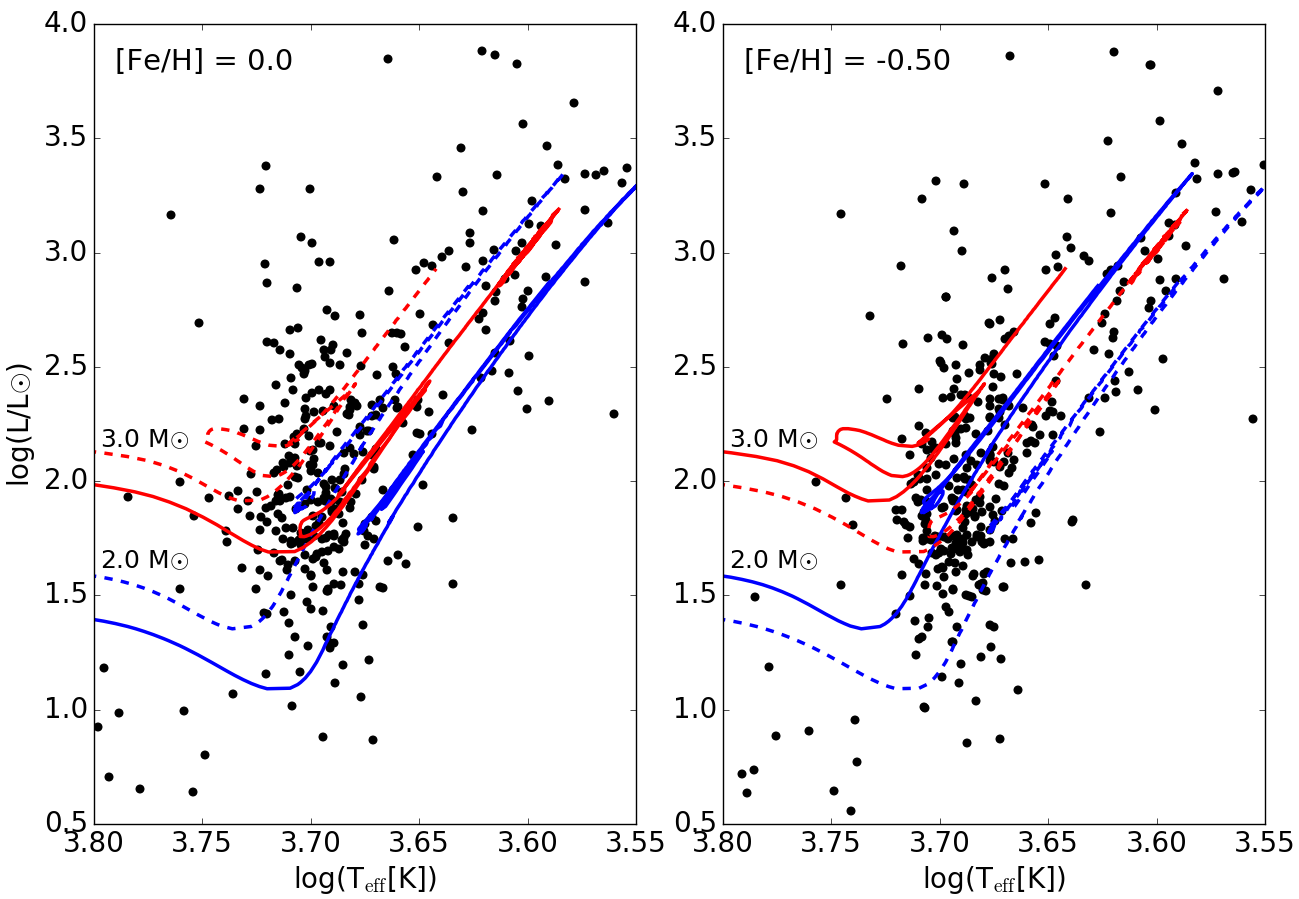}
\caption{\label{Fig:HRDFeH}
Same as Fig.~\ref{Fig:HRDStarevol} with the 2.0 and 3.0~\Msun\ (blue and red, respectively) evolutionary tracks (not covering the AGB) computed with [Fe/H] = 0.0 (left) and [Fe/H] = $-$0.5 (right). In each panel the solid tracks correspond to the indicated metallicity, which is also used to derive the atmospheric parameters of the targets; the dashed tracks correspond to the other metallicity ([Fe/H] = $-$0.5 and 0.0 in the left and right panel, respectively). 
}
\end{figure}

Evolutionary tracks for [Fe/H] = $-$0.25 have been superimposed for models of masses 1.0, 2.0, 3.0, 4.0, 5.0, and 6.0~\Msun, as labelled. Fig.~\ref{Fig:HRDFeH} shows the impact of metallicity on the 2.0~\Msun\ and 3.0~\Msun\ tracks (blue and red tracks, respectively). It appears that, for the RGB and the region occupied by the red clump, the [Fe/H] = $-$0.5 track of 2.0~\Msun\ covers the same region as the [Fe/H] = 0 track of 3.0~\Msun, leading to a strong degeneracy in the mass determination of a given star, which can only be lifted by knowing its metallicity.

\begin{figure}
\includegraphics[width=0.49\textwidth]{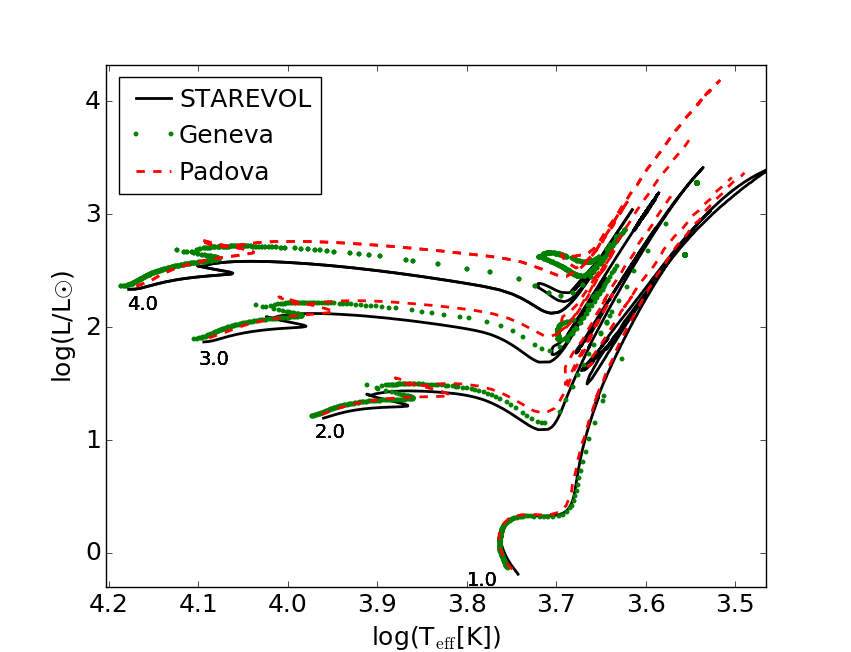}
\caption{\label{Fig:tracks}
Comparison among STAREVOL (solid black lines), Geneva (dotted green lines), and Padova (dashed red lines) evolutionary tracks up to the beginning of the AGB phase.
}
\end{figure}

Another important source of uncertainties is in the physics of the evolutionary models. Figure \ref{Fig:tracks} shows the difference among three sets of evolutionary tracks. The solid black tracks were computed with STAREVOL as described in Sect. \ref{Sect:STAREVOL} using the \cite{Asplund09} solar composition corresponding to Z$_{\odot}$\,=\,0.0134. The dotted green lines show the Geneva models from \cite{Ekstrom12} for solar composition as well, but with Z$_{\odot}$\,=\,0.014. Finally, the dashed red tracks correspond to the tracks from  \cite{Girardi00} for solar metallicity with Z$_{\odot}$\,=\,0.019. Apart from the scatter at the zero-age main sequence caused by the diverse solar metallicity values, varying descriptions of convection, overshooting, rotation, mass loss, etc., lead to uncertainties in the models in addition to our observational uncertainties. However, comparing evolutionary models is beyond the scope of this publication. Finally, barium stars are post-mass-transfer objects which, while on the main sequence, accreted mass from an AGB companion. This effect, which is not taken into account in the single-star evolutionary models that we use for comparison, could also affect their evolution.

A first interesting result apparent in the HRD of Fig.~\ref{Fig:HRDStarevol} is that several stars previously classified as Ba giants need to be reclassified as they are in fact dwarfs or subgiants. These are listed in Table ~\ref{Tab:Bad} and were selected from the criterion $L \le 10 $L$_{\odot}$.

We also report that most of subgiant CH stars (blue triangles) populate the region of the HRD associated with Ba dwarfs. Some of these stars, located around the 1.0~\Msun~track, seem to lie on the main sequence, contrary to what is implied by their designation as subgiants. This could have been guessed from the high gravities ($\log g \ge 4.0$) inferred by \citet{Luck-Bond-1991} in their spectroscopic analysis of subgiant CH stars. Other subgiant CH stars seem to be genuine subgiants, but they fall amidst stars classified as Ba dwarfs. The exact designation of these stars depends on the evolutionary models used for comparison, as the extension of the main sequence depends on the chosen overshoot parameter and the consideration or not of internal rotation. However, it seems clear that there is no clear separation between Ba dwarfs and CH subgiants in the HRD, so that these designations should not be taken literally.

\begin{table*}
\caption{\label{Tab:Bad}
List of dwarf barium stars, not previously recognized as such, from the catalogues of \citet{Lu1983}, \citet{Lu1991}, and \citet{Bartkevicius1996}.  
}
\begin{tabular}{llp{0.3\linewidth}p{0.3\linewidth}}
\hline
Name & TYC & Ref. & Rem.\\
\hline
HD 8270     &   8036-564-1 & \citet{Lu1991}, dwarf nature confirmed by \citet{Pereira2005}      \\
HD 13551        & 8851-37-1 & \citet{Lu1991}, dwarf nature confirmed by \citet{Pereira2005}\\
HD 22589     & 4722-19-1 & \citet{Lu1991}, dwarf nature confirmed by \citet{Pereira2005} & orbit available \\
CpD -44 5038 &  7735-447-1 & \citet{Lu1991}\\
BD -10 4311 & 5630-641-1 & \citet{Lu1991} & orbit available\\
HD 197481  & 7457-641-1 & \citet{Lu1991} \\
CS 22180-0013 &5279-303-1 & Bartkevicius\\
HIP 19050  & 1814-348-1 & Bartkevicius, classified as R\\
HD 175179       & 5123-323-1 & Bartkevicius PM*, moderate enhancement (0.2 -- 0.3~dex of Y and Ba) \citep{Bensby2014}\\
\hline
\end{tabular}
\end{table*}

A strong concentration of barium-rich stars is also found in the red clump, as is further discussed in Sect.~\ref{Sect:masses}.

\section{Mass distribution}
\label{Sect:masses}
 
We determine the masses from a comparison with stellar evolutionary tracks from the STAREVOL code. Three different metallicities have been considered: [Fe/H]$=-0.5, -0.25,$ and 0. To ensure self-consistency, when comparing the location of the target stars in the HRD with these tracks, we use the stellar parameters derived from the SED fit associated with the same metallicity (Sect.~\ref{Sect:atmosphere} and  Table~\ref{Tab:master}).

To derive stellar masses, we use a statistical approach that takes into account the time that a star spends at a given location in the  HRD. We first interpolated all the evolutionary tracks using a constant time step of 10,000 yrs. This interpolation thus produces more points in a long-lasting phase (such as the main sequence or the red clump). This allows us to fill the HRD with many points (around $17\times10^6$), each characterized by a value for $T_{\rm e}$, $L,$ and mass $m_{T,L}$; our method takes into account the mass lost by the star when ascending the RGB. For a given star, with a known temperature $T_o$ and luminosity $L_o$, along with their associated errors ($\sigma_{To}$, $\sigma_{Lo}$), we then estimate its mass, $m_o$, by computing a Gaussian average of the masses for all the points located in a restricted neighbourhood around $T_o, L_o$. More specifically, the mass of the star is given by 
 \begin{equation}
m_o =  \frac{1}{2 \pi \sigma_{To} \sigma_{Lo}} \sum_{T_e} \sum_L \frac{m_{T,L}^2}{\rho_{T,L}} e^{-\frac{(T_e-T_o)^2}{2 \sigma^2_{To}}}e^{-\frac{(L-L_o)^2}{2 \sigma^2_{Lo}}},
\end{equation}
where $\rho_{T,L}$ is a normalization factor given by
\begin{equation}
\rho_{T,L} = \frac{1}{2 \pi \sigma_{To} \sigma_{Lo}} \sum_{T_e} \sum_L m_{T,L} e^{-\frac{(T_e-T_o)^2}{2 \sigma^2_{To}}}e^{-\frac{(L-L_o)^2}{2 \sigma^2_{Lo}}}.
\end{equation}

\begin{figure}
\includegraphics[width=0.49\textwidth]{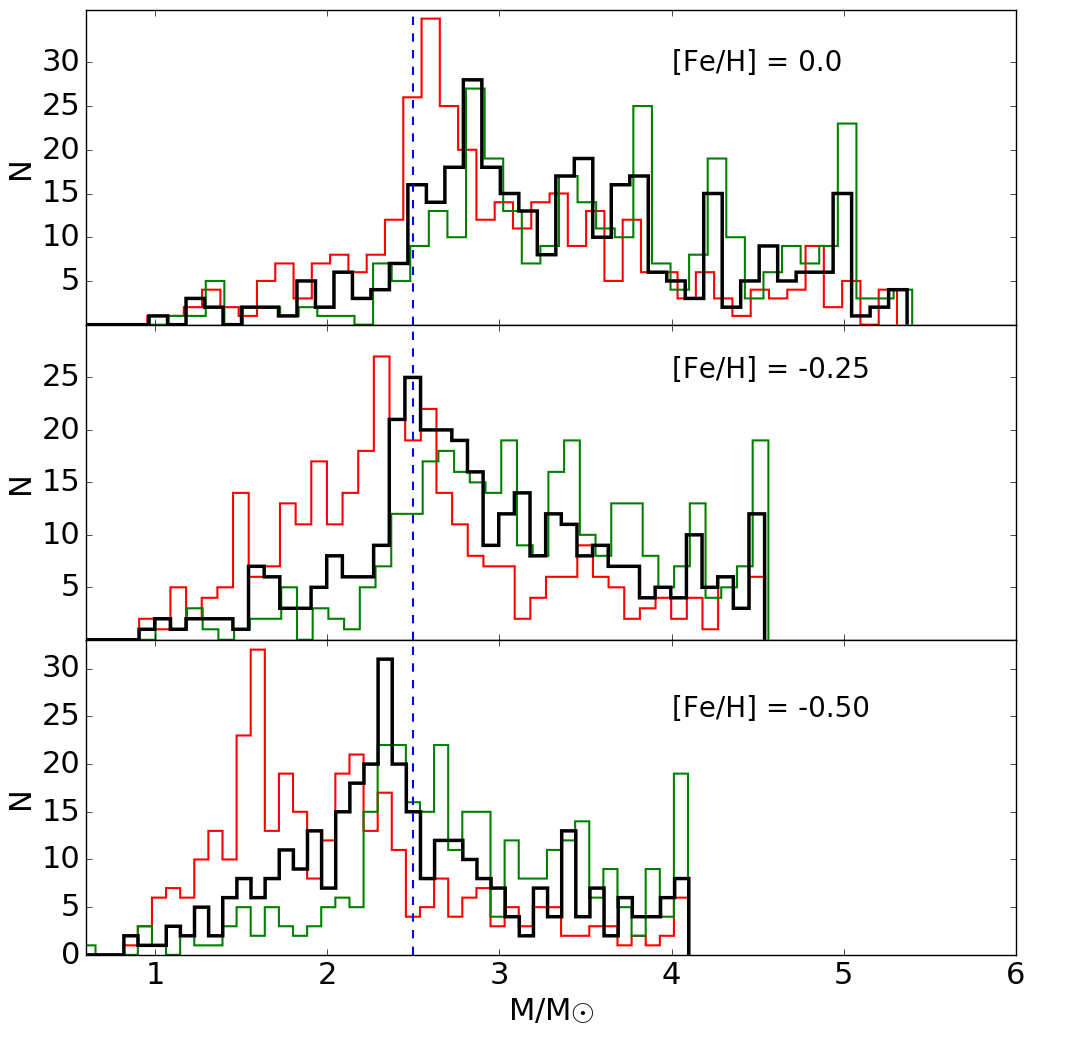}
\caption{\label{Fig:massesSTAREVOL}
Mass distribution for giant barium stars (thus excluding barium dwarfs, CH stars, and CH subgiants), from a comparison with the STAREVOL evolutionary tracks, for the three  metallicities [Fe/H]~$= 0$, [Fe/H]~$=-0.25$, and [Fe/H]~$=-0.5$. Adopting $T_{\rm eff} \pm \sigma_{\rm Teff}$ and $L \pm \sigma_L$, we determine the red (-) and green (+) histograms. The vertical dashed line indicates the peak of the [Fe/H]~$=-0.25$ distribution.
}
\end{figure}

We thereby derive the mass for all our objects, for the three computed metallicities. The distribution  for giant barium stars (thus excluding barium dwarfs, CH stars, and CH subgiants) is shown in Fig.~\ref{Fig:massesSTAREVOL} to highlight the sensitivity on the metallicity: if one adopts the solar metallicity for all the targets, the mass distribution exhibits a modest peak at 2.8~\Msun, and this peak is superimposed on a broader distribution extending from 2.5 to 5.5~\Msun. Adopting instead a lower metallicity [Fe/H]$= -0.5$ for all targets, the peak of the mass distribution is better marked and shifted towards lower masses (2.3~\Msun).  As shown in Fig.~\ref{Fig:FeH}, the metallicity distribution of barium stars peaks at [Fe/H]~$\sim$~$-$0.14. Hence, we conclude that the mass distribution obtained by adopting the average metallicity of -0.25 should be close to reality. In that case, the distribution can be described by two Gaussians: a main peak at 2.5~\Msun\ with a standard deviation of 0.18~\Msun\ and a broader tail at higher masses (up to 4.5~\Msun), which peaks at 3~\Msun\ with a standard deviation of 1~\Msun.

\begin{figure}
\includegraphics[width=0.49\textwidth]{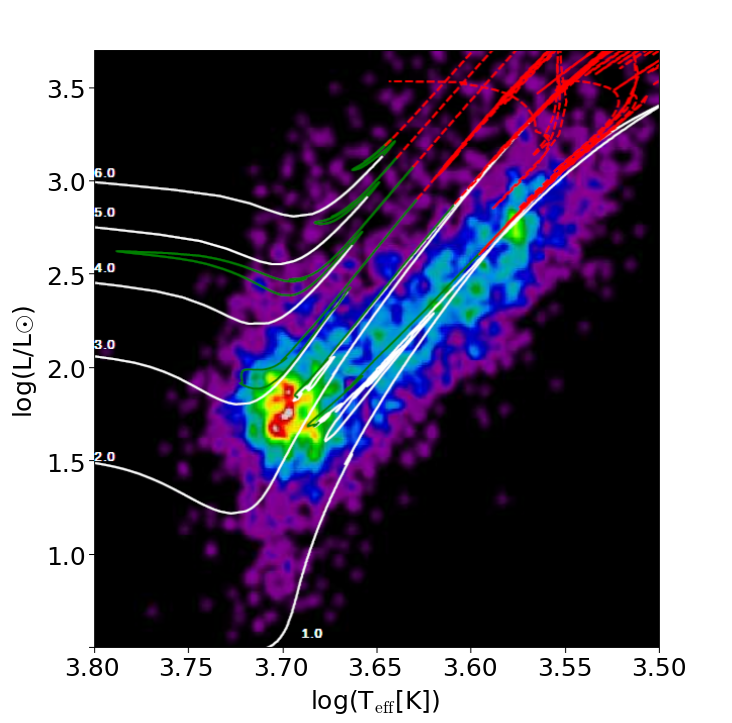}
\caption{\label{Fig:HRDFamaey}
Distribution of a comparison sample of K and M giants from \citet{Famaey-2005} in the HRD. Data are presented as a contour density plot. STAREVOL evolutionary tracks for stars with initial masses 1, 2, 3, 4, 5, and 6~\Msun\ (from  bottom to top) and [Fe/H]~=~-0.25 are superimposed; white colours correspond to Hertzsprung-gap and RGB evolution, green colours to core He-burning,  and red to AGB evolution. Stars from the Y group correspond to the 
faint extension of the stellar density in the regions covered by the 4, 5, and 6~\Msun\ evolutionary tracks. 
}
\end{figure}

\begin{figure}
\includegraphics[width=0.49\textwidth]{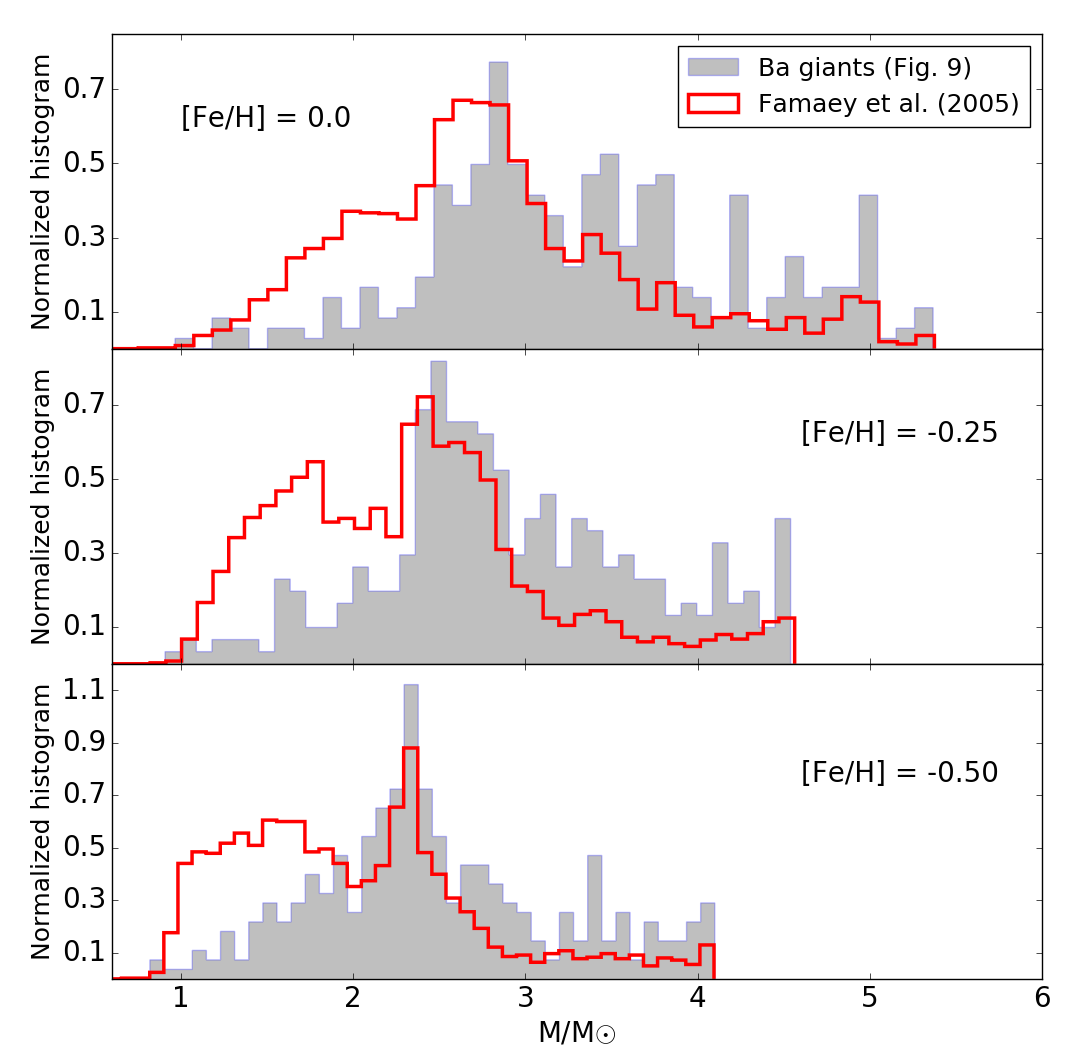}
\caption{\label{Fig:massPP44}
Same as Fig.~\ref{Fig:massesSTAREVOL} for the comparison sample of K and M giants (solid red line) from \citet{Famaey-2005}. The hatched histogram corresponds to barium stars. 
}
\end{figure}

\subsection{Comparison with M- and K-type giants}
We compare the mass distribution of barium giants with the sample of 5952 K and 739 M giants from \citet{Famaey-2005}.
These authors performed a Bayesian classification of this large sample into different kinematical groups, based not only on their kinematics (Tycho-2 proper motions and CORAVEL radial velocities; \citealt{Hog2000}; \citealt{Baranne-79}, respectively) but also on their luminosity, which is derived from a Bayesian estimate based on the Hipparcos parallax. The  B (`Background') group constitutes the smooth velocity ellipsoid, whereas the Y (`Young') group exhibits all signatures of young stars, i.e. small velocity dispersions and scale height. The average metallicity of stars belonging to the B group was estimated to be about $-0.2$ from Fig.~15 of \citet{Famaey-2005} and $-0.12\pm0.18$ from \cite{Girardi2001} as quoted by \citet{Famaey-2005}.
The close match between the average metallicities of barium giants (Fig.~\ref{Fig:FeH}) and the B group of comparison giants is encouraging, since this match indicates that B group giants are an appropriate sample to compare with the barium giants.
To locate B-group giants  in the HRD, it was necessary to use  various calibrations from \citet{Bessell-98}, notably ($V-I, V-K$) and ($V-K$, \Teff) to convert Hipparcos $V-I$ indices into temperatures, and ($V-K, BC_K$) to convert visual absolute magnitudes into luminosities. The resulting HRD is shown in Fig.~\ref{Fig:HRDFamaey} and indicates a strong concentration of solar-metallicity stars with masses in the range 2 -- 3~\Msun in the red clump. 

\subsection{Discussion}
The derived mass distribution (Fig.~\ref{Fig:massPP44}) reveals that the samples of field K and M and barium giants behave similarly in terms of mass, except for a deficit of low-mass stars ($M \la 2.0$~\Msun, the exact threshold value depending on metallicity) among barium stars.  In the mass distribution of the comparison sample, the Y group populates the high-mass tail, whereas the B group is associated with the lower mass population, including the strong peak  around 2.4~\Msun\ with a standard deviation of 0.26~\Msun\ (for [Fe/H] = $-$0.25), where stars residing in the red clump accumulate. \citet{VanderSwaelmen2017} have also found a peak around 2.3~\Msun~in the mass distribution of a sample of binary red giants in open clusters. The position of the peak should therefore not be considered as a distinctive feature of barium giants.

The deficit of low-mass giants ($M \la 2.0$~\Msun) could, however, be specific to Ba stars. One could think that it is caused by the longer evolutionary timescales needed to form a low-mass barium star whose companion could then also be of low mass. However, this argument does not hold because a star with $M \gtrsim 1.3$~\Msun~reaches the tip of the AGB within 5 Gyr and has time to pollute its companion and form a low-mass Ba star of solar metallicity. Additionally, low-mass Ba stars do not necessarily have low-mass companions. Neither can we claim that the deficit of low-mass barium stars is attributed to the absence of third dredge-up in low-mass ($M \lesssim$ 1.3~\Msun) AGB stars \citep{Karakas2016} because evidence for s-process enrichment in stars with such low masses has been suggested based on the analysis of post-AGB stars \citep{DeSmedt2015}. The most probable explanation for the deficit of low-mass Ba giants is that they reached a large radius at the tip of the RGB, resulting in a premature shrinkage of the orbit caused by a unstable Roche-lobe overflow (unless they had a very long orbital period). We elaborate on the method to determine $R_{\rm RGB tip}$ in Sect. \ref{Sect:P-HRD}.

\begin{figure}
\includegraphics[width=0.49\textwidth]{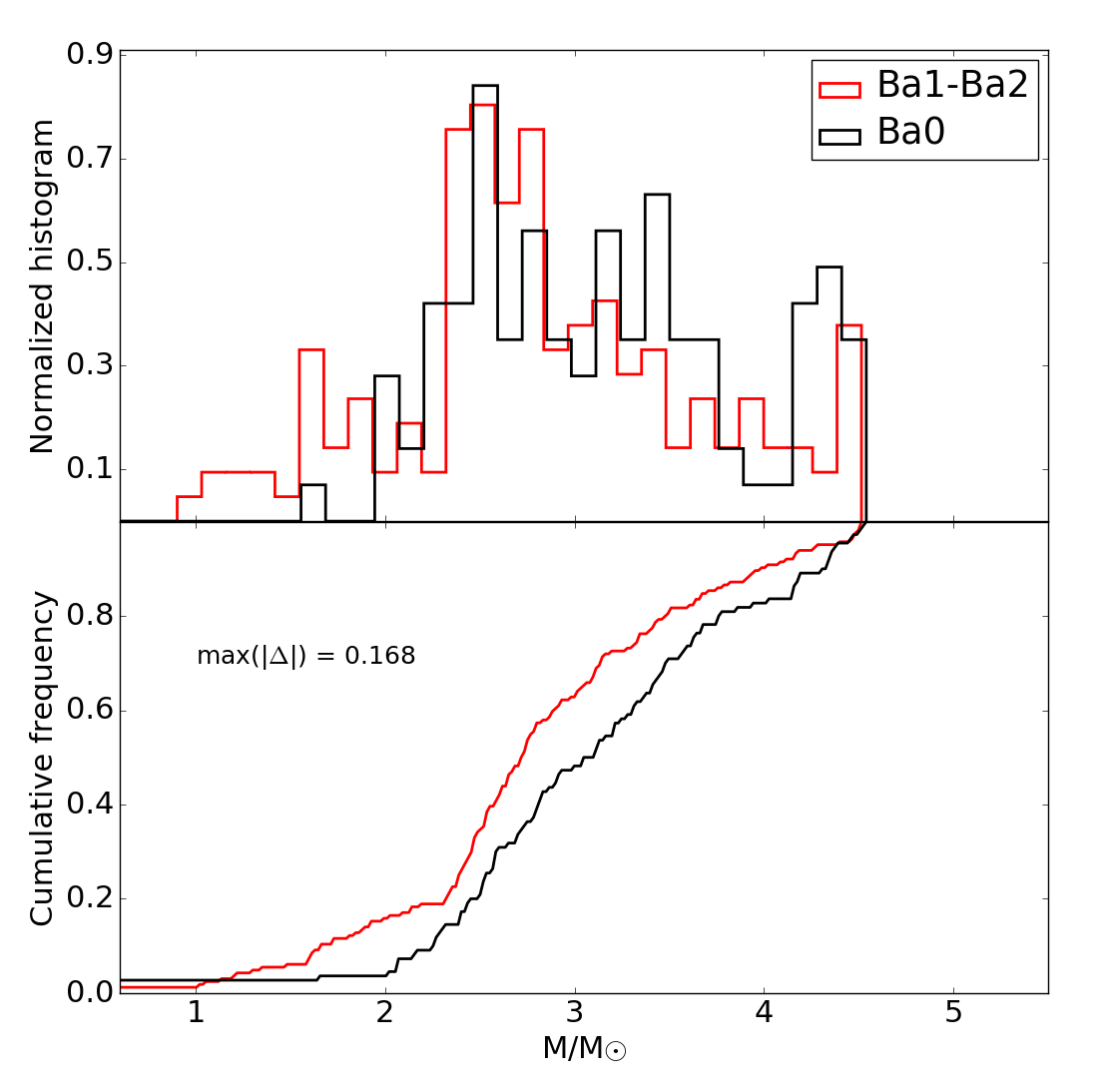}
\caption{\label{Fig:Bamm}
Comparison of the mass distributions for 164 mild (Ba1 -- Ba2) and 110 very mild (Ba0) barium stars (red and black curves, respectively) computed using evolutionary tracks with [Fe/H] = $-$0.25. Very mild barium stars are defined as having a \textit{Ba index} smaller than 1 in the catalogue of \citet{Lu1991}. Upper panel: Normalized histograms are shown. Lower panel: Cumulative frequency distributions are shown with the maximum absolute difference between these distributions amounting to 0.168, corresponding to a first-kind error of 13\% when rejecting the null hypothesis of identical distributions in a Kolmogorov-Smirnov test with an effective sample size of 48.2 [$= 164 \times 110 / (164 + 110)$]. 
}
\end{figure}

Among the barium-star sample, the high-mass tail is dominated by stars with a barium index, based on a visual inspection of the strength of the barium lines in the spectrum, lower than unity (named Ba0 in this paper), i.e. with a very moderate (if any) s-process enhancement (Fig.~\ref{Fig:Bamm}). Some of these stars, which appeared in the second edition (1991) of the barium-star catalogue of L\" u, could be bright giants, or even supergiants, where the barium lines are strengthened as a simple positive luminosity effect (i.e. an increase of the line strength with increasing luminosity) instead of being genuine barium stars. 

\begin{figure}
\includegraphics[width=0.49\textwidth]{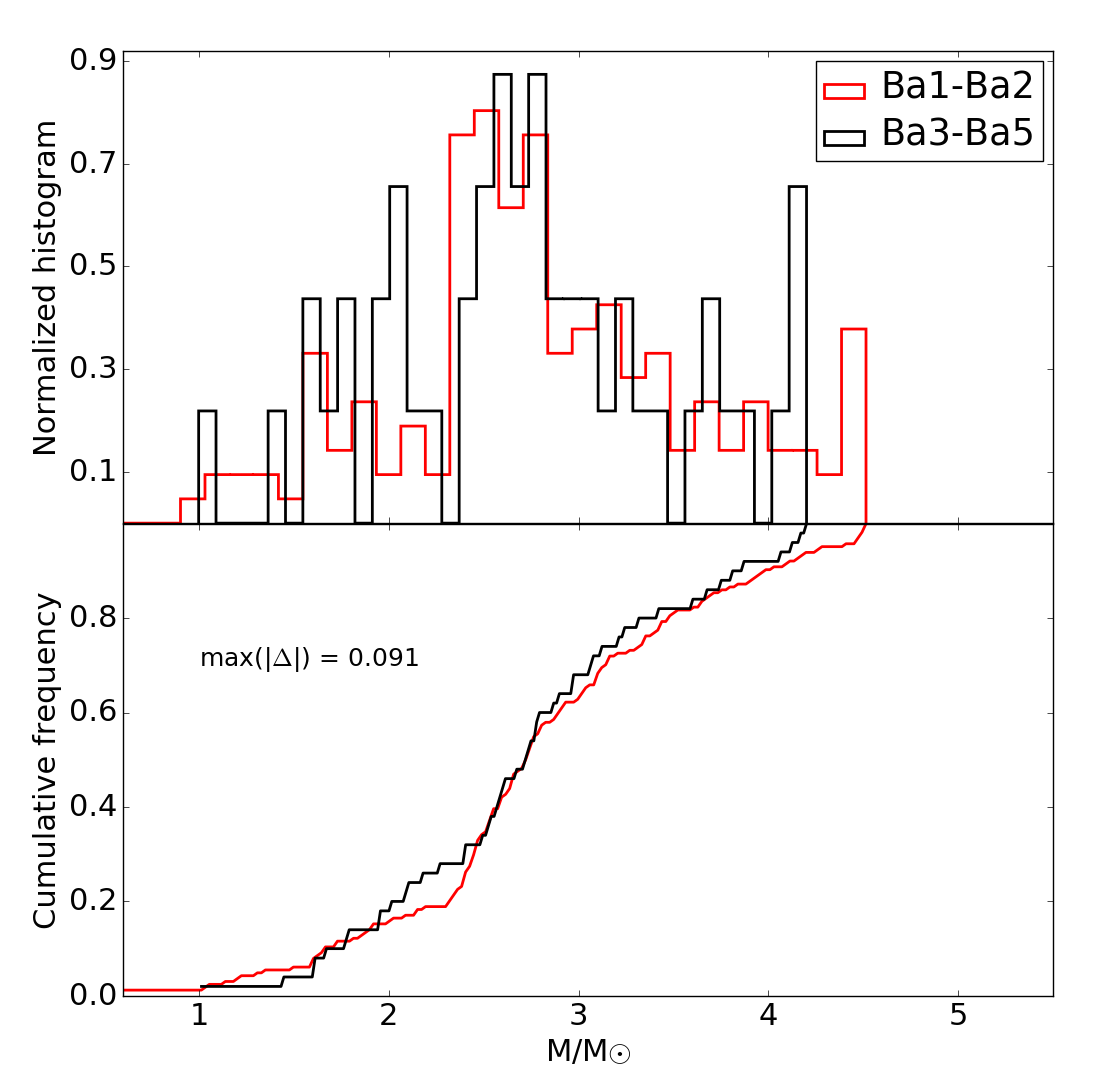}
\caption{\label{Fig:basm}
Comparison of the mass distributions for mild and strong barium stars (red and black curves, respectively) computed from evolutionary tracks for [Fe/H] = $-$0.25. Upper panel: Normalized histograms are shown. Lower panel: Cumulative frequency distributions with the maximum absolute difference between them amounting to 0.091.
}
\end{figure}

Finally, we re-investigate the statement by \citet{Mennessier1997} and \citet{Jorissen98}  that  mild (Ba1-Ba2) and  strong (Ba3-Ba5)
barium stars have somewhat different mass distributions, as they are characterized by masses in the range 2.5 -- 4.5~M$_\odot$
and 1 -- 3~M$_\odot$, respectively. As shown by Fig.~\ref{Fig:basm}, our analysis does not support this claim, since the mass distributions of 164 mild
and 50 strong barium giants are undistinguishable under the assumption that their average metallicities are similar. The maximum absolute difference between the two distributions is 0.091, which for an effective sample size of 38.3 ($= 164\times  50 / 214)$ corresponds to a very large first-kind error of 91\% for the Kolmogorov-Smirnov test.

\section{Location in the HRD and orbital period}
\label{Sect:P-HRD}

\begin{figure}
\includegraphics[width=0.49\textwidth]{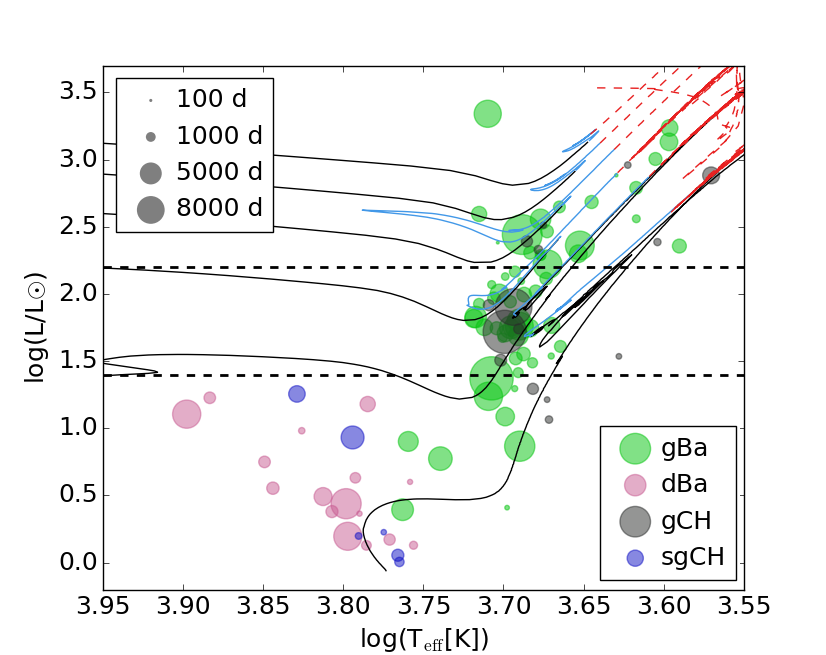}
\caption{\label{Fig:HR_P}
Same as Fig.~\ref{Fig:HRDStarevol}, but with the symbol size proportional to the orbital period.
}
\end{figure}

Fig.~\ref{Fig:HR_P} is a first attempt to correlate the location of a binary system in the HRD with its orbital period. We only include barium star systems for which the orbital period is known, and make the size of the symbols proportional to this value. We use the periods determined by \citet{Jorissen98} and Jorissen et al. (in preparation) for the giants (see also Table \ref{Tab:master}) and Escorza et al. (in preparation) for the main sequence and subgiant stars. We still keep the original labels they had in the queried catalogues and in Fig. \ref{Fig:HRDStarevol}.

\begin{figure}
\includegraphics[width=0.49\textwidth]{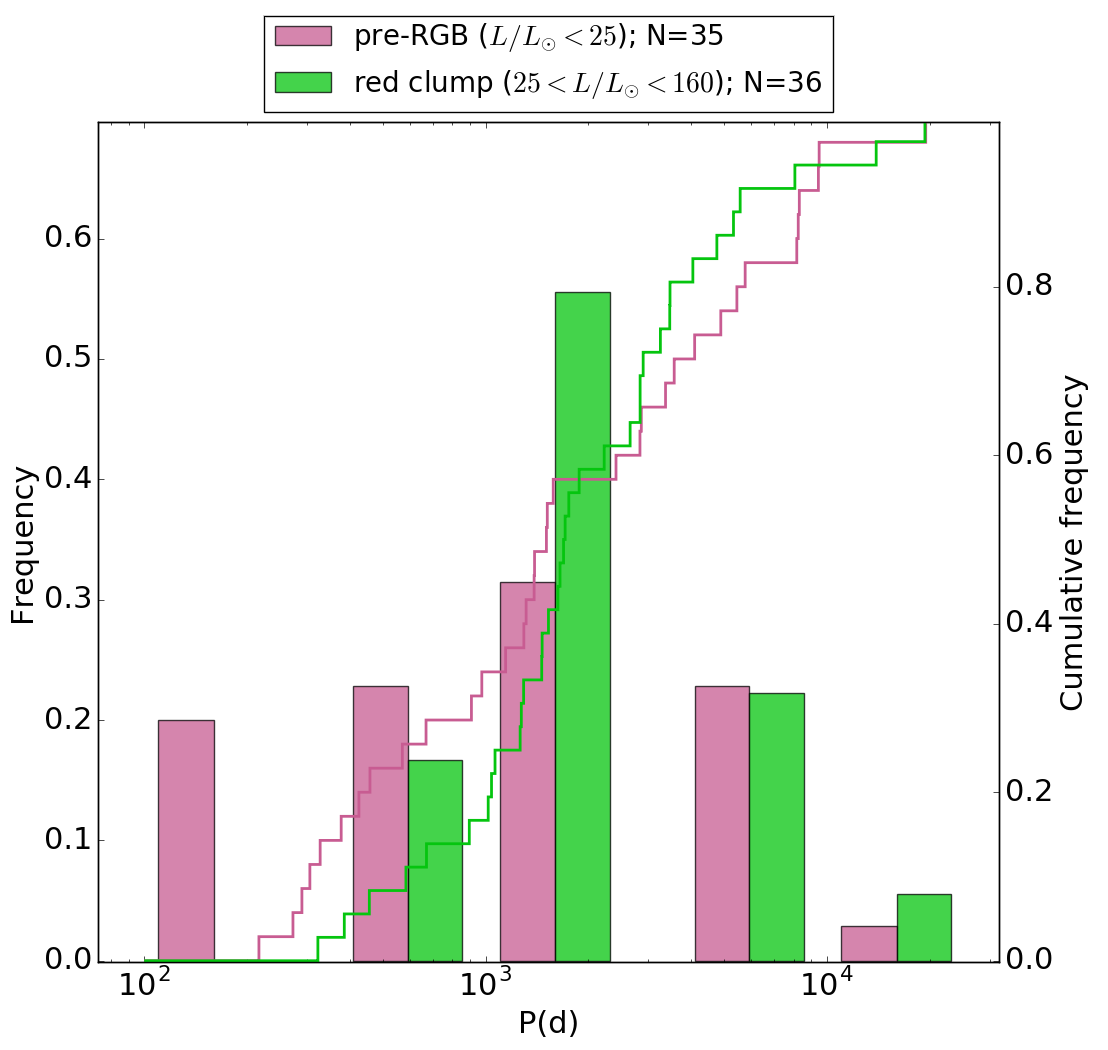}
\caption{\label{Fig:histo_P}
Normalized histogram (left scale, with bins of size 0.5 in log scale) and cumulative frequency distribution (right scale) for orbital 
periods of barium and related 
systems. Systems with $L/$L$_{\odot} \le 25$ are depicted in pink, and systems with $25 < L/$L$_{\odot} \le 160$) are depicted in green.}
\end{figure}

There seems to be a tendency for larger orbital periods in the red clump, as compared
to systems lying below (i.e. with luminosities from  the main sequence up to the red clump). In an attempt to quantify 
this effect, the sample has been split into two subsamples: one with $L/$L$_{\odot} \le 25$ ($\log L/$L$_{\odot} \le 1.4$) and the other with $25 < L/$L$_{\odot} \le 160$ ($1.5 < \log L/$L$_{\odot} \le 2.2$).
The first sample comprises main-sequence and subgiant stars whose current primary components never evolved up the RGB. The second sample, however, comprises systems in the red clump after their passage through the RGB tip, at least for systems with low-mass primaries, and a few less evolved objects, which happen to cross the red clump on their way to the RGB tip. Given the large radius reached at the RGB tip (see below), one may suspect that in the shorter period systems the giant filled its Roche lobe, and possibly evolved through a common-envelope stage into a cataclysmic variable or even a coalesced pair. Fig.~\ref{Fig:histo_P} confirms that very few systems (6/36 = 17\%) located in the red clump (green) have periods shorter than 1000~d and none of the systems have a period below 300~d. Compared to the less evolved systems (pink), there is indeed a deficit of short-period systems among red clump stars.

To apply the Kolmogorov-Smirnov (K-S) test, the maximum difference between the two cumulative frequency distributions was evaluated and amounts to $D = 0.2$, for an effective sample size of $N = 18 = m\times n/(m+n)$, where $m = 35$ and $n = 36$ are the sizes of the pre-RGB and red clump samples, respectively. The resulting effective difference $D N^{1/2} = 0.85$ corresponds to a first-risk error of 47\% to reject the null hypothesis incorrectly. Thus, the K-S test does not give us sufficient evidence to reject the null hypothesis and to conclude that the two distributions are different. However, it is known that the K-S test tends to be more sensitive near the centre of the distribution, while our samples seem to differ in the tails. Moreover, since the second-kind error has not been assessed because of the risk of incorrectly accepting the null hypothesis, the above result does not constitute a definite proof that the samples are extracted from the same parent population.

For this reason, we turned to another test ("four-box test") relying on the hypergeometric probability distribution. We divided our two samples (pre-clump and clump) into two subsamples with periods shorter and longer than 1000 days. Among the 35 pre-clump systems, 20 have short periods and 15 have long periods. Among the 36 ($\equiv N1$) systems in the clump, 30 have short periods and 6 have long periods. Hence, among the 71 ($\equiv N$) binaries, 21 binaries have short periods, so the probability of finding a short-period binary is $p = 21/71 = 0.3$. From this, we can now calculate the expected number of short-period objects among the clump sample as  $\tilde{x} = p \times N1 = 10.8$ instead of $6$ that are observed. In order to check whether or not this difference is  significant, we compute the variance on  $\tilde{x}$ as follows from the hypergeometric probability distribution ("small-sample statistics"): \\

$\sigma^{2}_{\tilde x} = N1 \times p \times (1-p) \times 
\frac{N-N1}{N-1}  = 3.78,$
i.e. $\sigma_{\tilde x} = 1.9$.\\

Hence, the expectation is  $\tilde{x} = 10.8 \pm 1.9$ or $8.9 < x < 12.7$ at 1 $\sigma$, meaning that the pre-clump and clump samples are different with a 2.5$\sigma$ significance.

We show in Table~\ref{Tab:RGB} that the 1000~d period corresponds to the period threshold below which a $\backsim$1\Msun~Ba dwarf fills its Roche lobe when it reaches the RGB tip, and that the less populated 400 -- 1000~d period range can be associated with the period cut-offs for stars in the mass range  [1.0 -- 2.5]~M$_{\odot}$. The critical periods for Roche-lobe overflow (RLOF) listed in Table~\ref{Tab:RGB} are derived as follows, assuming that the giant star fills its Roche lobe ($R_{\rm L_{1}}$) at the RGB tip (with radius $R_{\rm tip}$):
\begin{equation}
R_{L_{1}} = a\;\frac{0.49 q^{2/3}}{0.6 q^{2/3} + \ln( 1 + q^{1/3})} \equiv a\;f = R_{\rm tip}, 
\end{equation}
where $q = M_1/M_2$, with $M_1$ the mass of the giant star, and $M_2$ that of its WD companion. The critical semi-major axis of a circular orbit is then expressed as $a = R_{\rm L_{1}}/f = R_{\rm tip}/f$ and the corresponding period cut-off is
\begin{equation}
\label{Eq:Pcut}
P = \frac{(R_{\rm tip} / f)^{3/2}} {(M_1+M_2)^{1/2}}, 
\end{equation}
where $P$ is expressed in years, $R_{\rm tip}$ in AU (and is identified with the Roche radius $R_{\rm L_{1}}$ around star 1), and the component masses $M_{1,2}$ in \Msun.
The radius $R_{\rm tip}$ is given by the STAREVOL models. A direct comparison between this predictions and the observed values is not straightforward. The periods entering the above equation are pre-mass-transfer values, while the observed periods are the post-mass-transfer values. A detailed discussion of the relationship between initial and final periods is beyond the scope of this paper (see for example \citealt{Han95}, \citealt{Pols2003} or \citealt{Izzard2010}). Nevertheless, this analysis offers an order of magnitude estimate, although it should be considered with caution.

The absence of short-period barium stars in the red clump strongly suggests that RLOF on the RGB leads to systems that disappear from the barium star family. They could be the progenitors of short-period cataclysmic variables, or mergers. It is also tempting to associate these RGB-coalesced systems with early R-type carbon stars, which are another family of peculiar stars with a strong concentration in the red clump \citep{Knapp2001}, and none of these stars are  members of a binary system \citep{McClure1997}. The major problem with the hypothesis of a link between barium and R stars, however, is the lack of s-process element overabundances in early R-type carbon stars
\citep{Zamora2009} and of $^{12}$C in Ba stars. 

The segregation in terms of orbital periods between pre-red clump and red clump systems requires further discussion, because all these barium systems host a WD whose progenitor went through the thermally pulsing AGB (TP-AGB) phase. We should in principle expect a cut-off in the period distribution of the barium dwarf systems as well because AGB stars also reach very large radii.

\begin{figure}
\includegraphics[width=0.49\textwidth]{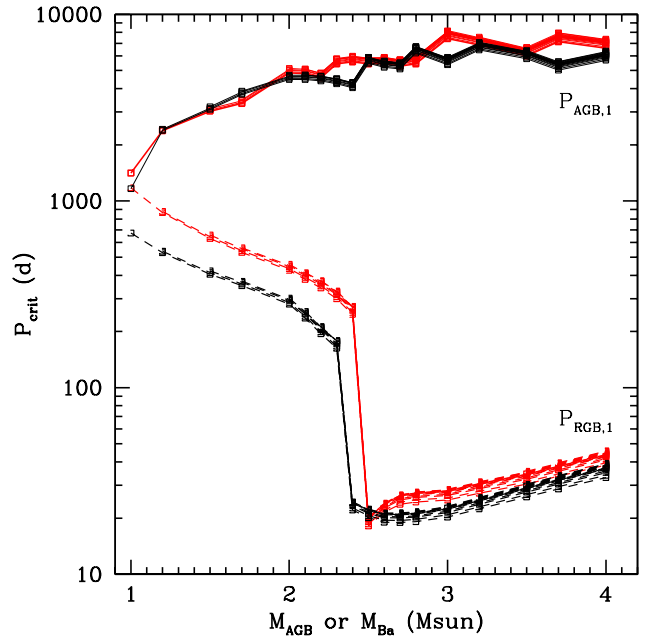}
\caption{\label{Fig:Pcrit}
Critical periods below which one of the components fills its Roche lobe at either the tip of the RGB ($P_{\rm RGB, 1}$, see text; dashed line at the bottom of the figure) or at the tip of the AGB ($P_{\rm AGB, 1}$, solid line at the top of the figure). The input data $R_{\rm RGB,tip}$ and $R_{\rm AGB,tip}$ are taken from the STAREVOL grid (red curves solar metallicity $Z = 0.0134$; black curves: [Fe/H]$ = -0.5$ or $Z = 0.0043$). Various superimposed lines in a series correspond to various barium-star masses. The corresponding $M_{\rm Ba}$ value may be identified from the starting point of the curve, which then extends towards increasing masses; for a given barium-star mass, the AGB component of mass $M_{\rm AGB}$ -- the former primary of the binary system -- is necessarily larger. 
The critical period $P_{\rm RGB, 2}$ (see text) does not strongly depend upon the mass of the WD companion. Therefore, its value can simply be read off the $P_{\rm RGB, 1}$ curve  (bold dashed line) at the corresponding $M_{\rm Ba}$ value, which determines the key value $R_{\rm RGB,tip}$. Although $P_{\rm RGB, 1}$ and $P_{\rm RGB, 2}$ are not strictly equal, they are 
not different enough to warrant a specific $P_{\rm RGB, 2}$ curve  that would jeopardize the clarity of the figure (see also Table~\ref{Tab:RGB}). This is a corrected version of the original Fig. 17.}
\end{figure}

In fact, three different critical periods must thus be considered: (i) $P_{\rm RGB, 1}$, when the initially more massive  component (now a WD) reaches the RGB tip, i.e. when in the above equations, $M_1 = M_{\rm 1@RGBtip}$ and $M_2 = M_{\rm Ba}$; (ii) $P_{\rm AGB, 1}$, when the initially more massive component reaches the AGB tip, i.e. when $M_1 = M_{\rm 1@AGBtip} = M_{\rm WD}$ and $M_2 = M_{\rm Ba}$; and (iii) $P_{\rm RGB, 2}$, when the initially less massive  component (the current barium star) reaches the RGB tip, its companion being a WD, i.e. when $M_1 = M_{\rm Ba}$ and $M_2 = M_{\rm WD}$. A supplementary condition is that the barium-star mass ($M_{\rm Ba}$) is lower than the initial mass of its companion, the WD progenitor; we neglected the mass transferred onto the barium star to make the computations easier. Periods    $P_{\rm RGB, 1}$ and $P_{\rm AGB, 1}$ are shown in Fig.~\ref{Fig:Pcrit}, whereas $P_{\rm RGB, 2}$ are listed in Table~\ref{Tab:RGB} (assuming $M_2 = M_{\rm WD}$ = 0.6~M$_{\odot}$). There is a weak dependence of $P_{\rm RGB, 1}$ on the initial mass ratio.

We clearly see that RLOF at the RGB tip occurs for all systems with a low-mass ($M \lesssim 2.2$\Msun) primary and periods shorter than 110~d (and shorter than 1000~d for 1~\Msun\ stars; see also Table~\ref{Tab:RGB}). This period threshold is consistent with the orbital-period segregation observed in the barium stars. On the other hand, intermediate-mass stars ($2.2 \lesssim M$/M$_{\odot} \lesssim 8 $) do not expand as much as their lower mass counterparts and the critical period is too short to affect the barium star population.

Another interesting conclusion may be drawn from the other (apparent) puzzle emerging from the present analysis; that is how the long-period thresholds (600 to 7500~d; $P_{\rm AGB, 1}$ in Fig.~\ref{Fig:Pcrit}) imposed by the passage of the WD progenitor through the AGB tip can be reconciled with the much shorter periods currently observed among barium systems. This is the long-standing puzzle identified by, for example \citet{Webbink1988}, \citet{Boffin-Jorissen-88}, and \citet{Pols2003}. Most barium systems should thus have gone through RLOF at the AGB tip, but this RLOF cannot have led to a catastrophic orbital shrinkage. In that case, we seek to understand why the  barium systems seemingly disappeared after RLOF occurring on the RGB, but not after RLOF occurring along the AGB of the former primary and current WD. The key difference between RLOF close to AGB and RGB tip is the mass of the residual envelope, which is generally much larger at the RGB tip. Hence, RLOF close to AGB tip does not involve a massive envelope, and we may expect a modest orbital shrinkage at that stage. Alternatively, if RLOF starts when the mass ratio is close to, or below, unity (a likely possibility with a stripped AGB star), a common evolution can be avoided, since the orbital separation expands soon after the beginning of mass transfer. This moderate shrinkage could then account for the fact that the observed periods for RGB systems shown in Fig.~\ref{Fig:histo_P} are shorter than the AGB threshold periods indicated in Fig.~\ref{Fig:Pcrit}. The possibility of intense mass loss and angular momentum loss on the AGB should also be considered (e.g. \citealt{Jahanara05}; \citealt{Boffin15}).

An interesting conclusion ensues by applying the  previous arguments to the early stage of the binary system, when the former primary ascended the RGB with the future barium star still on the main sequence. For such systems with an orbital period shorter than $P_{\rm RGB, 1}$,  RLOF with dramatic orbital shrinkage must have occurred, thus probably removing from the barium-star family those systems that had an initial period shorter than 1000~d with a primary mass of 1~M$_\odot$ or shorter than 200~d with a primary mass of 2.2~M$_\odot$. Stars in the mass range 1 -- 2.2~M$_\odot$ would have led to a WD with a mass below  0.58~M$_\odot$, according to the initial - final mass relationship inferred from the STAREVOL tracks. Therefore, we may expect a deficit of low-mass WDs among barium systems. This issue will be investigated in a forthcoming paper.

\begin{table}
\caption[]{\label{Tab:RGB} 
Red giant branch-tip radius and luminosity for stars of different masses, according to the STAREVOL evolutionary tracks for  $Z = 0.0134$, and the corresponding critical period $P_{\rm RGB, 2}$ (see text).   To be specific, the mass of the WD companion was fixed at  0.60~\Msun.}
\begin{tabular}{llllllll}
\hline
$M_{\rm now}$ & $M_{\rm initial}$ &$\log(L/$L$_{\odot})$  & $R_{\rm RGB tip}$ & $a$& $P$\\
(M$_{\odot}$) & (M$_{\odot}$) & & (R$_{\odot})$ &  (R$_{\odot}$)  & (d)\\
\hline
1.0 & 1.2 &  3.42 &  184 & 438 & 835 \\
1.5 & 1.6  & 3.43 &  162 & 352 & 524\\
2.0 &  2.1 & 3.39 &  136 & 283 & 340\\
2.5 &  2.6 & 3.64 &  190 & 380 & 484\\
3.0 & 3.0  & 3.21 &    92 & 177 & 141 \\
\hline
\end{tabular}
\end{table}

\section{Conclusions}

We have constructed an HRD of a sample of barium and related stars with unprecedented accuracy thanks to the TGAS parallaxes. From the comparison of the location of these stars on the HRD with STAREVOL evolutionary models, we find that the reclassification of some of these stars is needed. Several stars classified as Ba giants in the past are in fact dwarfs or subgiants. Additionally, there is no separation between subgiant CH stars and dwarf Ba in the HRD, contrary to what is implied by their designations.

The mass distribution derived for barium giants shows a deficit of low-mass stars ($M \la 2.0$~\Msun; the exact threshold depends on metallicity), as compared to normal K and M field giants. This deficit is probably a consequence of a premature shrinkage of the orbits with shorter periods, as less massive stars reach large radii on the tip of the RGB. Additionally, stars with a barium index below Ba1, introduced by \cite{Lu1991} in the second edition of his catalogue, populate the high-mass tail of the distribution and probably are not genuine Ba stars but bright giants where the Ba lines are strengthened by a positive luminosity effect. Finally, and in contrast to previous claims, we do not see any difference between the mass distributions of mild (Ba1 -- Ba2) and strong (Ba3 -- Ba5) barium stars, at least when adopting the same average metallicity [Fe/H] = $-$0.25 for both groups. It seems that there is no correlation between Ba abundance and Ba stellar mass.

There is a deficit of barium systems with short periods among those populating the red clump, as a result of the period threshold set by the large radii reached at the tip of the RGB. The systems missing from the red clump population could have coalesced as a result of unstable RLOF close to the tip of the RGB.

The accuracy of the TGAS parallaxes marks a significant step forward for the derivation of barium star masses. However, to obtain information on individual stars, the derivation of metallicities from spectroscopic data for each target is crucial. A consistent derivation of the metallicities would also allow us to conclude if the apparent difference between the metallicity distributions of dwarf and giant barium stars is real.

\begin{acknowledgements}
This research has been funded by the Belgian Science Policy Office under contract BR/143/A2/STARLAB and the Fonds voor Wetenschappelijk
Onderzoek Vlaanderen (FWO). A.E. acknowledges support from the FWO. S.V.E acknowledges the Fondation ULB. L.S and D.P. are Senior FNRS Research Associates. D.K acknowledges support from BELSPO. The authors want to acknowledge the referee, Claudio B. Pereira, for his interest and very valuable input. This work has made use of data from the European Space Agency (ESA) mission \textit{Gaia} (\url{https://www.cosmos.esa.int/gaia}), processed by the \textit{Gaia} Data Processing and Analysis Consortium (DPAC, \url{https://www.cosmos.esa.int/web/gaia/dpac/consortium}). Funding for the DPAC has been provided by national institutions, in particular the institutions participating in the \textit{Gaia} Multilateral Agreement. This research has also made use of the SIMBAD database, operated at CDS, Strasbourg, France.
\end{acknowledgements}

\bibliographystyle{aa}
\bibliography{articles}

\begin{thebibliography}{65}
\expandafter\ifx\csname natexlab\endcsname\relax\def\natexlab#1{#1}\fi

\bibitem[{{Asplund} {et~al.}(2009){Asplund}, {Grevesse}, {Sauval}, \&
  {Scott}}]{Asplund09}
{Asplund}, M., {Grevesse}, N., {Sauval}, A.~J., \& {Scott}, P. 2009, \araa, 47,
  481

\bibitem[{{Baranne} {et~al.}(1979){Baranne}, {Mayor}, \& {Poncet}}]{Baranne-79}
{Baranne}, A., {Mayor}, M., \& {Poncet}, J.~L. 1979, Vistas in Astronomy, 23,
  279

\bibitem[{{Bartkevi\~cius}(1996)}]{Bartkevicius1996}
{Bartkevi\~cius}, A. 1996, Baltic Astronomy, 5, 217

\bibitem[{{Bensby} \& {Feltzing}(2006)}]{Bensby2006}
{Bensby}, T. \& {Feltzing}, S. 2006, \mnras, 367, 1181

\bibitem[{{Bensby} {et~al.}(2014){Bensby}, {Feltzing}, \& {Oey}}]{Bensby2014}
{Bensby}, T., {Feltzing}, S., \& {Oey}, M.~S. 2014, \aap, 562, A71

\bibitem[{{Bessell} {et~al.}(1998){Bessell}, {Castelli}, \&
  {Plez}}]{Bessell-98}
{Bessell}, M.~S., {Castelli}, F., \& {Plez}, B. 1998, \aap, 333, 231

\bibitem[{{Bidelman} \& {Keenan}(1951)}]{Bidelman1951}
{Bidelman}, W.~P. \& {Keenan}, P.~C. 1951, \apj, 114, 473

\bibitem[{{Boffin}(2015)}]{Boffin15}
{Boffin}, H.~M.~J. 2015, {Mass Transfer by Stellar Wind}, ed. H.~M.~J.
  {Boffin}, G.~{Carraro}, \& G.~{Beccari}, 153

\bibitem[{{Boffin} \& {Jorissen}(1988)}]{Boffin-Jorissen-88}
{Boffin}, H. M.~J. \& {Jorissen}, A. 1988, \aap, 205, 155

\bibitem[{{Bond} {et~al.}(2008){Bond}, {Lauretta}, {Tinney}, {Butler}, {Marcy},
  {Jones}, {Carter}, {O'Toole}, \& {Bailey}}]{Bond2008}
{Bond}, J.~C., {Lauretta}, D.~S., {Tinney}, C.~G., {et~al.} 2008, \apj, 682,
  1234

\bibitem[{{de Castro} {et~al.}(2016){de Castro}, {Pereira}, {Roig}, {Jilinski},
  {Drake}, {Chavero}, \& {Sales Silva}}]{deCastro16}
{de Castro}, D.~B., {Pereira}, C.~B., {Roig}, F., {et~al.} 2016, \mnras, 459,
  4299

\bibitem[{{De Smedt} {et~al.}(2015){De Smedt}, {Van Winckel}, {Kamath}, \&
  {Wood}}]{DeSmedt2015}
{De Smedt}, K., {Van Winckel}, H., {Kamath}, D., \& {Wood}, P.~R. 2015, \aap,
  583, A56

\bibitem[{{Degroote} {et~al.}(2011){Degroote}, {Acke}, {Samadi}, {Aerts},
  {Kurtz}, {Noels}, {Miglio}, {Montalb{\'a}n}, {Bloemen}, {Baglin}, {Baudin},
  {Catala}, {Michel}, \& {Auvergne}}]{Degroote2011}
{Degroote}, P., {Acke}, B., {Samadi}, R., {et~al.} 2011, \aap, 536, A82

\bibitem[{{Edvardsson} {et~al.}(1993){Edvardsson}, {Andersen}, {Gustafsson},
  {Lambert}, {Nissen}, \& {Tomkin}}]{Edvardsson1993}
{Edvardsson}, B., {Andersen}, J., {Gustafsson}, B., {et~al.} 1993, \aap, 275,
  101

\bibitem[{{Ekstr{\"o}m} {et~al.}(2012){Ekstr{\"o}m}, {Georgy}, {Eggenberger},
  {Meynet}, {Mowlavi}, {Wyttenbach}, {Granada}, {Decressin}, {Hirschi},
  {Frischknecht}, {Charbonnel}, \& {Maeder}}]{Ekstrom12}
{Ekstr{\"o}m}, S., {Georgy}, C., {Eggenberger}, P., {et~al.} 2012, \aap, 537,
  A146

\bibitem[{ESA(1997)}]{ESA1997}
ESA. 1997, The Hipparcos and Tycho Catalogues (ESA)

\bibitem[{{Famaey} {et~al.}(2005){Famaey}, {Jorissen}, {Luri}, {Mayor}, {Udry},
  {Dejonghe}, \& {Turon}}]{Famaey-2005}
{Famaey}, B., {Jorissen}, A., {Luri}, X., {et~al.} 2005, \aap, 430, 165

\bibitem[{{Ferguson} {et~al.}(2005){Ferguson}, {Alexander}, {Allard}, {Barman},
  {Bodnarik}, {Hauschildt}, {Heffner-Wong}, \& {Tamanai}}]{Ferguson05}
{Ferguson}, J.~W., {Alexander}, D.~R., {Allard}, F., {et~al.} 2005, \apj, 623,
  585

\bibitem[{{Gaia Collaboration} {et~al.}(2016{\natexlab{a}}){Gaia
  Collaboration}, {Brown}, {Vallenari}, {Prusti}, {de Bruijne}, {Mignard},
  {Drimmel}, {Babusiaux}, {Bailer-Jones}, {Bastian}, \&
  et~al.}]{Gaia-collaborationDR1}
{Gaia Collaboration}, {Brown}, A.~G.~A., {Vallenari}, A., {et~al.}
  2016{\natexlab{a}}, \aap, 595, A2

\bibitem[{{Gaia Collaboration} {et~al.}(2016{\natexlab{b}}){Gaia
  Collaboration}, {Prusti}, {de Bruijne}, {Brown}, {Vallenari}, {Babusiaux},
  {Bailer-Jones}, {Bastian}, {Biermann}, {Evans}, \& et~al.}]{GaiaMission}
{Gaia Collaboration}, {Prusti}, T., {de Bruijne}, J.~H.~J., {et~al.}
  2016{\natexlab{b}}, \aap, 595, A1

\bibitem[{{Girardi} {et~al.}(2000){Girardi}, {Bressan}, {Bertelli}, \&
  {Chiosi}}]{Girardi00}
{Girardi}, L., {Bressan}, A., {Bertelli}, G., \& {Chiosi}, C. 2000, \aaps, 141,
  371

\bibitem[{{Girardi} \& {Salaris}(2001)}]{Girardi2001}
{Girardi}, L. \& {Salaris}, M. 2001, \mnras, 323, 109

\bibitem[{{Gontcharov}(2012)}]{Gontcharov2012}
{Gontcharov}, G.~A. 2012, Astronomy Letters, 38, 87

\bibitem[{{Gustafsson} {et~al.}(2008){Gustafsson}, {Edvardsson}, {Eriksson},
  {J{\o}rgensen}, {Nordlund}, \& {Plez}}]{Gustafsson2008}
{Gustafsson}, B., {Edvardsson}, B., {Eriksson}, K., {et~al.} 2008, \aap, 486,
  951

\bibitem[{{Han} {et~al.}(1995){Han}, {Eggleton}, {Podsiadlowski}, \&
  {Tout}}]{Han95}
{Han}, Z., {Eggleton}, P.~P., {Podsiadlowski}, P., \& {Tout}, C.~A. 1995,
  \mnras, 277, 1443

\bibitem[{{Herwig} {et~al.}(1997){Herwig}, {Bloecker}, {Schoenberner}, \& {El
  Eid}}]{Herwig97}
{Herwig}, F., {Bloecker}, T., {Schoenberner}, D., \& {El Eid}, M. 1997, \aap,
  324, L81

\bibitem[{{H{\o}g} {et~al.}(2000){H{\o}g}, {Fabricius}, {Makarov}, {Urban},
  {Corbin}, {Wycoff}, {Bastian}, {Schwekendiek}, \& {Wicenec}}]{Hog2000}
{H{\o}g}, E., {Fabricius}, C., {Makarov}, V.~V., {et~al.} 2000, \aap, 355, L27

\bibitem[{{Iglesias} \& {Rogers}(1996)}]{IglesiasRogers96}
{Iglesias}, C.~A. \& {Rogers}, F.~J. 1996, \apj, 464, 943

\bibitem[{{Izzard} {et~al.}(2010){Izzard}, {Dermine}, \& {Church}}]{Izzard2010}
{Izzard}, R.~G., {Dermine}, T., \& {Church}, R.~P. 2010, \aap, 523, A10

\bibitem[{{Jahanara} {et~al.}(2005){Jahanara}, {Mitsumoto}, {Oka}, {Matsuda},
  {Hachisu}, \& {Boffin}}]{Jahanara05}
{Jahanara}, B., {Mitsumoto}, M., {Oka}, K., {et~al.} 2005, \aap, 441, 589

\bibitem[{{Jorissen} \& {Boffin}(1992)}]{JorissenBoffin92}
{Jorissen}, A. \& {Boffin}, H.~M.~J. 1992, in Binaries as Tracers of Star
  Formation, ed. A.~{Duquennoy} \& M.~{Mayor}, 110--131

\bibitem[{{Jorissen} {et~al.}(1998){Jorissen}, {Van Eck}, {Mayor}, \&
  {Udry}}]{Jorissen98}
{Jorissen}, A., {Van Eck}, S., {Mayor}, M., \& {Udry}, S. 1998, \aap, 332, 877

\bibitem[{{K{\"a}ppeler} {et~al.}(2011){K{\"a}ppeler}, {Gallino}, {Bisterzo},
  \& {Aoki}}]{Kappeler2011}
{K{\"a}ppeler}, F., {Gallino}, R., {Bisterzo}, S., \& {Aoki}, W. 2011, Reviews
  of Modern Physics, 83, 157

\bibitem[{{Karakas} \& {Lugaro}(2016)}]{Karakas2016}
{Karakas}, A.~I. \& {Lugaro}, M. 2016, \apj, 825, 26

\bibitem[{{Keenan}(1942)}]{Keenan1942}
{Keenan}, P.~C. 1942, \apj, 96, 101

\bibitem[{{Knapp} {et~al.}(2001){Knapp}, {Pourbaix}, \& {Jorissen}}]{Knapp2001}
{Knapp}, G., {Pourbaix}, D., \& {Jorissen}, A. 2001, \aap, 371, 222

\bibitem[{{Lindegren} {et~al.}(2016){Lindegren}, {Lammers}, {Bastian},
  {Hern{\'a}ndez}, {Klioner}, {Hobbs}, {Bombrun}, {Michalik}, {Ramos-Lerate},
  {Butkevich}, {Comoretto}, {Joliet}, {Holl}, {Hutton}, {Parsons},
  {Steidelm{\"u}ller}, {Abbas}, {Altmann}, {Andrei}, {Anton}, {Bach},
  {Barache}, {Becciani}, {Berthier}, {Bianchi}, {Biermann}, {Bouquillon},
  {Bourda}, {Br{\"u}semeister}, {Bucciarelli}, {Busonero}, {Carlucci},
  {Casta{\~n}eda}, {Charlot}, {Clotet}, {Crosta}, {Davidson}, {de Felice},
  {Drimmel}, {Fabricius}, {Fienga}, {Figueras}, {Fraile}, {Gai}, {Garralda},
  {Geyer}, {Gonz{\'a}lez-Vidal}, {Guerra}, {Hambly}, {Hauser}, {Jordan},
  {Lattanzi}, {Lenhardt}, {Liao}, {L{\"o}ffler}, {McMillan}, {Mignard}, {Mora},
  {Morbidelli}, {Portell}, {Riva}, {Sarasso}, {Serraller}, {Siddiqui}, {Smart},
  {Spagna}, {Stampa}, {Steele}, {Taris}, {Torra}, {van Reeven}, {Vecchiato},
  {Zschocke}, {de Bruijne}, {Gracia}, {Raison}, {Lister}, {Marchant},
  {Messineo}, {Soffel}, {Osorio}, {de Torres}, \& {O'Mullane}}]{Lindegren16}
{Lindegren}, L., {Lammers}, U., {Bastian}, U., {et~al.} 2016, \aap, 595, A4

\bibitem[{{L\"u}(1991)}]{Lu1991}
{L\"u}, P.~K. 1991, \aj, 101, 2229

\bibitem[{{L\"u} {et~al.}(1983){L\"u}, {Dawson}, {Upgren}, \& {Weis}}]{Lu1983}
{L\"u}, P.~K., {Dawson}, D.~W., {Upgren}, A.~R., \& {Weis}, E.~W. 1983, \apjs,
  52, 169

\bibitem[{{Luck} \& {Bond}(1991)}]{Luck-Bond-1991}
{Luck}, R.~E. \& {Bond}, H.~E. 1991, \apjs, 77, 515

\bibitem[{{Luri} \& {Arenou}(1997)}]{Luri-1997}
{Luri}, X. \& {Arenou}, F. 1997, in ESA Special Publication, Vol. 402,
  Hipparcos - Venice '97, ed. R.~M. {Bonnet}, E.~{H{\o}g}, P.~L. {Bernacca},
  L.~{Emiliani}, A.~{Blaauw}, C.~{Turon}, J.~{Kovalevsky}, L.~{Lindegren},
  H.~{Hassan}, M.~{Bouffard}, B.~{Strim}, D.~{Heger}, M.~A.~C. {Perryman}, \&
  L.~{Woltjer}, 449--452

\bibitem[{{Marigo}(2002)}]{Marigo02}
{Marigo}, P. 2002, \aap, 387, 507

\bibitem[{{McClure}(1997)}]{McClure1997}
{McClure}, R.~D. 1997, \pasp, 109, 256

\bibitem[{{McClure} {et~al.}(1980){McClure}, {Fletcher}, \&
  {Nemec}}]{McClure1980}
{McClure}, R.~D., {Fletcher}, J.~M., \& {Nemec}, J.~M. 1980, \apj, 238, L35

\bibitem[{{McClure} \& {Woodsworth}(1990)}]{McClure1990}
{McClure}, R.~D. \& {Woodsworth}, A.~W. 1990, ApJ, 352, 709

\bibitem[{{Mennessier} {et~al.}(1997){Mennessier}, {Luri}, {Figueras}, {Gomez},
  {Grenier}, {Torra}, \& {North}}]{Mennessier1997}
{Mennessier}, M.~O., {Luri}, X., {Figueras}, F., {et~al.} 1997, \aap, 326, 722

\bibitem[{{Michalik} {et~al.}(2015){Michalik}, {Lindegren}, \&
  {Hobbs}}]{Michalik2015}
{Michalik}, D., {Lindegren}, L., \& {Hobbs}, D. 2015, \aap, 574, A115

\bibitem[{{North} {et~al.}(1994){North}, {Berthet}, \& {Lanz}}]{North1994}
{North}, P., {Berthet}, S., \& {Lanz}, T. 1994, \aap, 281, 775

\bibitem[{{North} {et~al.}(2000){North}, {Jorissen}, \& {Mayor}}]{North00}
{North}, P., {Jorissen}, A., \& {Mayor}, M. 2000, in IAU Symposium, Vol. 177,
  The Carbon Star Phenomenon, ed. R.~F. {Wing}, 269

\bibitem[{{Pereira}(2005)}]{Pereira2005}
{Pereira}, C.~B. 2005, \aj, 129, 2469

\bibitem[{{Pols} {et~al.}(2003){Pols}, {Karakas}, {Lattanzio}, \&
  {Tout}}]{Pols2003}
{Pols}, O.~R., {Karakas}, A.~I., {Lattanzio}, J.~C., \& {Tout}, C.~A. 2003, in
  Astronomical Society of the Pacific Conference Series, Vol. 303, Symbiotic
  Stars Probing Stellar Evolution, ed. R.~L.~M. {Corradi}, J.~{Mikolajewska},
  \& T.~J. {Mahoney}, 290

\bibitem[{{Pourbaix} \& {Jorissen}(2000)}]{Pourbaix2000}
{Pourbaix}, D. \& {Jorissen}, A. 2000, \aaps, 145, 161

\bibitem[{{Schr{\"o}der} \& {Cuntz}(2007)}]{SchroderCuntz07}
{Schr{\"o}der}, K.-P. \& {Cuntz}, M. 2007, \aap, 465, 593

\bibitem[{{Siess}(2006)}]{Siess2006}
{Siess}, L. 2006, \aap, 448, 717

\bibitem[{{Siess} \& {Arnould}(2008)}]{SiessArnould08}
{Siess}, L. \& {Arnould}, M. 2008, \aap, 489, 395

\bibitem[{{Siess} {et~al.}(2000){Siess}, {Dufour}, \& {Forestini}}]{Siess00}
{Siess}, L., {Dufour}, E., \& {Forestini}, M. 2000, \aap, 358, 593

\bibitem[{{Skrutskie} {et~al.}(2006){Skrutskie}, {Cutri}, {Stiening},
  {Weinberg}, {Schneider}, {Carpenter}, {Beichman}, {Capps}, {Chester},
  {Elias}, {Huchra}, {Liebert}, {Lonsdale}, {Monet}, {Price}, {Seitzer},
  {Jarrett}, {Kirkpatrick}, {Gizis}, {Howard}, {Evans}, {Fowler}, {Fullmer},
  {Hurt}, {Light}, {Kopan}, {Marsh}, {McCallon}, {Tam}, {Van Dyk}, \&
  {Wheelock}}]{2MASS}
{Skrutskie}, M.~F., {Cutri}, R.~M., {Stiening}, R., {et~al.} 2006, \aj, 131,
  1163

\bibitem[{{Soubiran} {et~al.}(2016){Soubiran}, {Le Campion}, {Brouillet}, \&
  {Chemin}}]{PASTEL16}
{Soubiran}, C., {Le Campion}, J.-F., {Brouillet}, N., \& {Chemin}, L. 2016,
  \aap, 591, A118

\bibitem[{{Van der Swaelmen} {et~al.}(2017){Van der Swaelmen}, {Boffin},
  {Jorissen}, \& {Van Eck}}]{VanderSwaelmen2017}
{Van der Swaelmen}, M., {Boffin}, H.~M.~J., {Jorissen}, A., \& {Van Eck}, S.
  2017, \aap, 597, A68

\bibitem[{{Vassiliadis} \& {Wood}(1993)}]{VassiliadisWood93}
{Vassiliadis}, E. \& {Wood}, P.~R. 1993, \apj, 413, 641

\bibitem[{{Warner}(1965)}]{Warner1965}
{Warner}, B. 1965, \mnras, 129, 263

\bibitem[{{Webbink}(1988)}]{Webbink1988}
{Webbink}, R.~F. 1988, in Critical Observations Versus Physical Models for
  Close Binary Systems, ed. K.-C. {Leung}, 403--446

\bibitem[{{Weingartner} \& {Draine}(2001)}]{Weingartner2001}
{Weingartner}, J.~C. \& {Draine}, B.~T. 2001, \apj, 548, 296

\bibitem[{{Wenger} {et~al.}(2000){Wenger}, {Ochsenbein}, {Egret}, {Dubois},
  {Bonnarel}, {Borde}, {Genova}, {Jasniewicz}, {Lalo{\"e}}, {Lesteven}, \&
  {Monier}}]{SIMBAD}
{Wenger}, M., {Ochsenbein}, F., {Egret}, D., {et~al.} 2000, \aaps, 143, 9

\bibitem[{{Zamora} {et~al.}(2009){Zamora}, {Abia}, {Plez}, {Dom{\'{\i}}nguez},
  \& {Cristallo}}]{Zamora2009}
{Zamora}, O., {Abia}, C., {Plez}, B., {Dom{\'{\i}}nguez}, I., \& {Cristallo},
  S. 2009, \aap, 508, 909

\end{thebibliography}

\begin{appendix}
\section{Sample of barium and related stars with accurate parallaxes}

Table~\ref{Tab:master} lists the barium and related stars with good-precision Gaia and Hipparcos parallaxes (i.e. $\varpi / \sigma(\varpi) > 3$). The full list is available at CDS, Strasbourg.

\begin{sidewaystable*}
\caption[]{\label{Tab:master}
List of barium and related stars with good-quality Gaia and Hipparcos parallaxes (i.e. $\varpi / \sigma(\varpi) > 3$), and their atmospheric and orbital parameters: $P$ is the orbital period and $e$ the eccentricity (from Jorissen et al., in preparation), type is Ba3-5 (giant barium star with strong anomaly), Ba1-2 (giant barium star with mild anomaly), Ba0 (giant barium star with very mild anomaly), Bad (dwarf barium star), sgCH (subgiant CH), and CH.  [Fe/H] is the metallicity and 'Ref.' the corresponding reference (N/A if not available), $\varpi\pm\sigma$ and source are the parallax$\pm$ its error and source (Gaia, Hipparcos or Pourbaix \& Jorissen 2000). The following columns contain the effective temperature and luminosity, with their 1$\sigma$ ranges, for [Fe/H] = 0, [Fe/H] = $-$0.25, and [Fe/H] = $-$0.5~dex (see text).
}
\begin{tabular}{rrrrrrrrrrrrrrrrrrrrrrr}
\hline\\
Name & \multicolumn{1}{c}{$P$ }& \multicolumn{1}{c}{$e$} &      type    & [Fe/H] & \multicolumn{1}{c}{Ref.} &     \multicolumn{1}{c}{$\varpi\pm\sigma$}& source  & [Fe/H] & $T_{\rm eff, min}$ &\multicolumn{1}{c}{$T_{\rm eff}$} &$T_{\rm eff, max}$ & \multicolumn{1}{c}{$L_{\rm min}$} & \multicolumn{1}{c}{$L$} &\multicolumn{1}{c}{$L_{\rm max}$}\\
         & \multicolumn{1}{c}{(d)} &        &           &             &      &  \multicolumn{1}{c}{(mas)}  & && \multicolumn{1}{c}{(K)} & \multicolumn{1}{c}{(K)} & \multicolumn{1}{c}{(K)}&\multicolumn{1}{c}{(L$_\odot$)} & \multicolumn{1}{c}{(L$_\odot$)} &\multicolumn{1}{c}{(L$_\odot$)}  \\
\hline\\
HD 26   & 19364 & 0.08  & CH    & -0.45 & 1986ApJ...303..226S   & $3.40\pm0.30$ &       Gaia 
                                                                     & 0      & 5136     & 5290 & 5342&  52 &    61 &     74     &       \\
         & &        &           &             &      &   &  & -0.25& 4898       & 5003 & 5193&    44 &    52 & 63 & \\
         & &        &           &             &      &   &  & -0.5  & 4773 &  4901 &       5077 &  42 &    49 & 59 \\
HD 218   & - &  - &     Ba3-5   & N/A &         N/A     & $1.71\pm0.27$ &       Gaia    
                                                                    & 0       & 3787 &   3979 & 4058 &   507 &   680 &960 \\
         & &        &           &             &      &   &  & -0.25& 3782 &       3958 & 4045 &   511 &   686 &969 \\
         & &        &           &             &      &   &  & -0.5 &  3768 &       3943 & 4035 &   508 &   681 &963 \\
HD 749   & - &  - & Ba1-2       & -0.06& 2006A\&A...454..895 &$7.11\pm1.08$ &               Hipparcos       &
                                                                          0&    4687 &4909       & 4967  & 14.0 &        18.5 & 25.8\\
         & &        &           &             &      &   &  & -0.25&4657&       4892    & 4979    & 13.8 &        18.3 & 25.4\\
         & &        &           &             &      &   &  & -0.5  &4662&4795        & 4990    & 12.9 &   17.0        & 23.7\\                
HD 4084  & - &  - &Ba3-5&       -0.42   & 2016MNRAS..459...4299 &       $1.81\pm0.25$&          Gaia    &       
                                                                               0&4631 &4874&     4970&   126&    164&    221&\\
         & &        &           &             &      &   &  & -0.25&4609  &4894 &        4992&     127&  165&    223\\
         & &        &           &             &      &   &  & -0.5  &4574  &4941 &   4991&     128&        166&     224\\                         
HD 5322 & - &   - &Ba1-2&       -0.18   & 2016MNRAS..459...4299 &       $1.43\pm0.26$ &         Gaia    &       
                                                                             0 &4827   &5083&    5326       &497&        696 &1046 \\
         & &        &           &             &      &   &  & -0.25&4832        &5065&  5237    &487&   683 &1026\\
         & &        &           &             &      &   &  & -0.5  &4858       &4957&  5224    &457&   641     &964\\
HD 5395 & - &   - &     Ba1-2&  -0.24   & 2016MNRAS..440..1095 &       $15.84\pm       0.58$   &Hipparcos      &       
                                                                               0&5139    &5201   &5263 & 76&     82      &88\\
         & &        &           &             &      &   &  & -0.25&4903  &5081      &5178 &  77&        83    &89\\
         & &        &           &             &      &   &  & -0.5  &4660  &4836  &5087&  70&  76 &82\\           
HD 5424 &1881&  0.23    &dBa1-2&        -0.51 & 2006A\&A...454..895 &   $2.50\pm        0.27$ &               Gaia    &       
                                                                              0&4727&4888&       5015&   29&  36&        45\\
         & &        &           &             &      &   &  & -0.25&4667&4870&  4987&   29&     36&     45\\    
         & &        &           &             &      &   &  &  -0.5&4575&       4804&   4953&   29&  35& 44 \\      
...\\                   
\hline\\
\end{tabular}
\end{sidewaystable*}

\end{appendix}
\end{document}